\documentclass[submission, Phys]{SciPost}
\usepackage[T1]{fontenc}
\usepackage[utf8]{inputenc} 

\makeatletter
\newcommand{\subalign}[1]{%
  \vcenter{%
    \Let@ \restore@math@cr \default@tag
    \baselineskip\fontdimen10 \scriptfont\tw@
    \advance\baselineskip\fontdimen12 \scriptfont\tw@
    \lineskip\thr@@\fontdimen8 \scriptfont\thr@@
    \lineskiplimit\lineskip
    \ialign{\hfil$\m@th\scriptstyle##$&$\m@th\scriptstyle{}##$\crcr
      #1\crcr
    }%
  }
}

\newcommand{\be}{\begin{equation}}
\newcommand{\ee}{\end{equation}}
\newcommand{\bea}{\begin{eqnarray}}
\newcommand{\eea}{\end{eqnarray}}

\let\polishl\l  

\def\a{\alpha}
\def\b{\beta}
\def\g{\gamma}
\def\G{\Gamma}
\def\d{\delta}

\def\e{\epsilon}

\def\th{\theta}

\def\k{\kappa}
\def\l{\lambda}

\def\m{\mu}
\def\n{\nu}
\def\c{\xi}

\def\p{\pi}
\def\P{\Pi}
\def\r{\rho}
\def\s{\sigma}
\def\S{\Sigma}

\def\w{\omega}
\def\W{\Omega}

\def\blk{{\mathbf k}}

\def\blp{{\mathbf p}}
\def\blQ{{\mathbf q}}
\def\blr{{\mathbf r}}

\def\blA{{\mathbf A}}

\def\blQ{{\mathbf Q}}

\def\blU{{\mathbf U}}

\def\blA{{\mathbf A}}

\def\callG{\mbox{$\mathcal{G}$}}

\def\callN{\mbox{$\mathcal{N}$}}

\def\callT{\mbox{$\mathcal{T}$}}

\def\de{\partial}
\def\iif{\infty}
\def\bra{\langle}
\def\ket{\rangle}

\def\Im{{\rm Im}}

\def\1op{\hat{\mathbbm{1}}}
\def\nn{\nonumber}

\def\bz{\mathbf 0}

\hypersetup{
    colorlinks,
    linkcolor={red!50!black},
    citecolor={blue!50!black},
    urlcolor={blue!80!black}
}

\usepackage[bitstream-charter]{mathdesign}
\DeclareSymbolFont{usualmathcal}{OMS}{cmsy}{m}{n}
\DeclareSymbolFontAlphabet{\mathcal}{usualmathcal}

\begin{document}

\begin{center}{\Large \textbf{
Excitonic Bloch equations from first principles 
}}\end{center}

\begin{center}
Gianluca Stefanucci\textsuperscript{1,2},
and
Enrico Perfetto\textsuperscript{1,2}
\end{center}

\begin{center}
{\bf 1} Dipartimento di Fisica, Universit{\`a} di Roma Tor Vergata, Via della Ricerca Scientifica 1,
00133 Rome, Italy
\\
{\bf 2} INFN, Sezione di Roma Tor Vergata, Via della Ricerca Scientifica 1, 00133 Rome, Italy
\\
${}^\star$ {\small \sf gianluca.stefanucci@roma2.infn}
\end{center}

\begin{center}
\today
\end{center}


\section*{Abstract}
{\bf
The ultrafast conversion of coherent  excitons into incoherent excitons, 
as well as the subsequent exciton diffusion and thermalization, are 
central topics in current scientific research due to their relevance 
in optoelectronics, photovoltaics and photocatalysis.
Current approaches to the exciton dynamics 
rely on {\em model} Hamiltonians that depend on 
already screened electron-electron and electron-phonon couplings. 
In this work, we subject the state-of-the-art methods to scrutiny 
using the {\em ab initio} Hamiltonian for electrons and phonons. 
We offer a rigorous and intuitive proof demonstrating 
that the exciton dynamics governed by model Hamiltonians is affected by an overscreening of 
the electron-phonon interaction.      
The introduction of an auxiliary exciton 
species, termed the irreducible exciton, enables us to formulate a 
theory free from overscreening and derive the excitonic Bloch 
equations. These equations describe the time-evolution of coherent, 
irreducible, and incoherent excitons during and after the optical excitation.    
They are applicable beyond the linear regime,
and predict that the total number of excitons is preserved
when the external fields are switched off.  

}

\vspace{10pt}
\noindent\rule{\textwidth}{1pt}
\tableofcontents\thispagestyle{fancy}
\noindent\rule{\textwidth}{1pt}
\vspace{10pt}

\section{Introduction}
\label{sec:intro}

Transition-metal dichalcogenides, layered heterostructures, and other 
two-dimensional materials hold great promise for next-generation 
optoelectronic devices due to their rich excitonic 
landscape~\cite{wang_ultrafast_2015,wang_colloquium_2018,ma_recent_2019,zhai_2D_2020,cheng_recent_2019}.
The weak dielectric screening favors the existence (and coexistence) 
of well defined optically bright zero-momentum singlet 
excitons~\cite{he_tightly_2014,chernikov_exciton_2014,hill_observation_2015} as well as 
dark inter-valley and triplet 
excitons~\cite{moody_intrinsic_2015,palummo_exciton_2015,wu_exciton_2015,selig_exciton_2016}. 
Understanding the fundamental laws governing the formation and 
scattering of excitons 
is crucial for guiding researchers and accelerating progress.  
The momentum-resolved 
exciton dynamics can be monitored using time-resolved 
photoemission techniques~\cite{saile_resonant_1980,Weinelt_PhysRevLett.92.126801,Varene_ultrafast_2012,zhu_photoemission_2015,Deinert_ultrafast_2014,madeo_directly_2020,tanimura_time_2020,man_experimental_2021,dong_direct_2021,wallauer_momentum-resolved_2021,lin_exciton-driven_2022,perfetto_first-principles_2016,steinhoff_exciton_2017,rustagi_photoemission_2018}.
Laser pulses with subgap frequencies excite particle-hole pairs and 
transfer the light's coherence to them. During and shortly after the 
optical excitation, the nonequilibrium state features coherent (or 
virtual) particle-hole pairs bound by the Coulomb attraction, i.e.,
bright coherent excitons. For not too large excitation densities, this superfluid 
phase~\cite{ostreich_non-stationary_1993,hannewald_excitonic_2000,perfetto_pump-driven_2019,perfetto_time-resolved_2020,chan_giant_2023}
is stable only at clamped 
nuclei~\cite{perfetto_real_2022}. 
The charge imbalance created by the laser pulse sets the 
nuclear lattice in motion, and the macroscopic number of lattice modes 
(or phonons) is responsible for destroying exciton 
coherence~\cite{madrid_phonon-induced_2009,nie_ultrafast_2014,dey_optical_2016,sangalli_an-ab-initio_2018,stefanucci_from-carriers_2021,perfetto_real-time_2023} and 
diffusing excitons~\cite{bertoni_generation_2016,uddin_neutral_2020,perfetto_ultrafast_2021}. 
Bright coherent excitons are then converted into 
bright and dark incoherent excitons, whose energy and momentum can be 
inferred from the measured time-resolved and angle-resolved photocurrent.

The complex dynamics of coherent and incoherent excitons coupled to 
phonons has become a central focus in current scientific research.   
A straightforward strategy to deal with the problem consists of using 
model bosonic Hamiltonians for excitons and phonons. 
Built on the pioneering works by Toyozawa and 
Hopfield~\cite{toyozawa_theory_1958,hopfield_on-the-energy_1961,toyozawa_interband_1964},
this approach dates back to the 
early 
sixties~\cite{usui_excitations_1960,segall_phonon-assisted_1968,haug_electron_1984}. 
Nonetheless, it faces 
important conceptual issues that remain to be 
addressed~\cite{combescot_why_2005,combescot_exciton-exciton_2007,paleari_exciton-phonon_2022}.    
The major issue pertains with the fact that
model Hamiltonians are written in terms of 
screened electron-electron ($e$-$e$) and electron-phonon 
($e$-$ph$) interactions. This leads to double counting 
and overscreening already at the leading order in perturbation 
theory. Techniques based on the cluster 
expansion~\cite{kira_many-body_2006,axt_nonlinear_1998,thranhardt_quantum_2000,brem_exciton_2018,christiansen_theory_2019} 
suffer from the same conceptual drawbacks, as the underlying Hamiltonian 
is affected by the same problems, i.e., a
screened $e$-$e$ and $e$-$ph$ coupling. The equivalence between the two approaches 
becomes evident when 
expanding the $e$-$ph$ interaction in terms of an 
interaction between (multiple) 
particle-holes and 
phonons~\cite{ivanov_self-consistent_1993,katsch_theory_2018}, which, 
to the lowest 
order, is equivalent to a 
bosonization.

Despite the conceptual issues, the equations of motion for the 
coupled dynamics of coherent and incoherent 
excitons~\cite{axt_nonlinear_1998,thranhardt_quantum_2000,jankovic_dynamics_2015,brem_exciton_2018,lengers_phonon-mediated_2020} 
have a great appeal, and have been extensively used in the literature. 
The fundamental question we address in this work 
is: Can these equations be derived from the ab initio Hamiltonian of electrons 
and phonons~\cite{stefanucci_in-and-out_2023}? 
In this study, we provide a conclusive negative answer, pinpointing the 
issue as rooted in the overscreening of the $e$-$ph$ 
interaction, and derive the correct result.  
Our derivation is 
based on Green's function many-body theory, where screening, 
quasi-particle renormalization, and phonon frequencies emerge 
naturally from the diagrammatic treatment of the ab initio 
Hamiltonian of electrons and phonons. 
The developed theory contains correlation effects up to 
second-order in the excitation density and treats excitons and 
particle-hole pairs (plasma) on equal footing.

The paper is organized as follows. In Section~\ref{prelsec} we 
introduce the connection between excitons and two-particle
Green's function, and briefly go through the ab initio 
Hamiltonian for electrons and phonons. We also present key 
properties of the Bethe-Salpeter equation for systems with $e$-$e$ 
and $e$-$ph$ interactions. In Sections~\ref{cohxsec} 
and~\ref{incexsec} we discuss the minimal approximation to the self-energy and 
exchange-correlation kernel leading to a coupled dynamics of 
coherent and incoherent excitons. The inclusion of the 
inelastic exciton-phonon scattering is elaborated in 
Section~\ref{xxsec}. The final outcome, summarized in 
Section~\ref{SBEEsec}, is the excitonic Bloch  equations --
a comprehensive set of coupled equations for coherent, 
irreducible and incoherent excitons. Finally, in Section~\ref{consec} 
we provide a summary of our main findings and outline future avenues 
for research and development. 

\section{Preliminaries}
\label{prelsec}

\subsection{Excitons and two-particle Green's function}
\label{aihamsec}

We consider a semiconductor with quasiparticle energies 
$\e_{c\blk}$ for conduction electrons in band $c$ with momentum 
$\blk$, and $\e_{v\blk}$ for valence 
electrons in band $v$ with momentum 
$\blk$. The band indices $c,v$ are spin-orbital indices. 
If the system Hamiltonian is 
invariant under spin rotations then $c=(a,\s)$, $v=(b,\s)$ can be 
written as pairs of indices, the first describing the orbital part 
and the second describing the spin-projection onto, say, the $z$ axis.
In the quasi-particle basis the Coulomb  
amplitude for the scattering process $(\m\blk+\blQ\,,\n'\blk') \to 
(\m'\blk'+\blQ\,,\n \blk)$ reads
\begin{align}
v_{\m\blk+\blQ\,\n'\blk'\,\n\blk\;\m'\blk'+\blQ}=\bra
\m\blk+\blQ\,\n'\blk'|\frac{1}{|\hat{\blr}-\hat{\blr}'|}
|\n \blk\;\m'\blk'+\blQ\ket.
\end{align}
We normalize the quasiparticle wavefunctions to unity so that the 
Coulomb amplitude scales like $1/\callN_{\blk}$,  $\callN_{\blk}$ 
being the number of $\blk$-points in the first Brillouin zone.
The exciton is 
a bound electron-hole  pair of the crystal with clamped nuclei, 
whose energy and eigenfunction 
satisfy the eigenvalue equation
\begin{align}
E^{\blQ}_{cv\blk}A^{\l\blQ}_{cv\blk}-
\sum_{c'v'\blk'}
K^{\rm HSEX,\blQ}_{cv\blk,c'v'\blk'}A^{\l\blQ}_{c'v'\blk'}=
E_{\l\blQ}A^{\l\blQ}_{cv\blk},
\label{eigeneqX}
\end{align}
where $E^{\blQ}_{cv\blk}=\e_{c\blk+\blQ}-\e_{v\blk}$ and 
\begin{align}
K^{\rm HSEX,\blQ}_{cv\blk,c'v'\blk'}=
W_{c\blk+\blQ\,v'\blk'\,v\blk\,c'\blk'+\blQ}-
v_{c\blk+\blQ\,v'\blk'\,c'\blk'+\blQ\,v\blk}=
W^{\blQ}_{cv\blk,c'v'\blk'}-v_{c\blk+\blQ\,v'\blk'\,c'\blk'+\blQ\,v\blk}
\label{hsex}
\end{align}
is the irreducible Hartree plus screened exchange (HSEX) kernel, $W$ being the statically 
screened interaction~\cite{Rohlfing_PRL1998,onida_electronic_2002}.
At any fixed $\blQ$ the exciton wavefunctions are chosen orthonormal, 
i.e., 
$\sum_{cv\blk}A^{\l\blQ\ast}_{cv\blk}A^{\l'\blQ}_{cv\blk}=\d_{\l\l'}$.
Henceforth we use the word excitons for all solutions (bound and 
unbound) of Eq.~(\ref{eigeneqX}). Thus, excitons and particle-hole 
pairs (plasma) are treated on equal footing.

We define the 
exciton creation operator
\begin{align}
\hat{X}^{\dag}_{\l\blQ}=\sum_{cv\blk}A^{\l \blQ}_{cv\blk}\,
\hat{d}^{\dag}_{c\blk+\blQ}\hat{d}_{v\blk},
\end{align}
where the operators $\hat{d}_{\m\blk}$ annihilate an electron of 
momentum $\blk$ in band $\m$, and satisfy the anticommutation 
relations 
$\{\hat{d}_{\m\blk},\hat{d}^{\dag}_{\m'\blk'}\}=\d_{\m\m'}\d_{\blk,\blk'}$. 
The  number operator for excitons of 
type $\l$ and momentum $\blQ$ is then
\begin{align}
\hat{N}_{\l \blQ}=\hat{X}^{\dag}_{\l\blQ}\hat{X}_{\l\blQ}.
\end{align}	
Looking at Eq.~(\ref{eigeneqX}) it 
is tempting to construct an effective boson Hamiltonian for excitons, 
i.e., 
$\hat{H}^{\rm x}=\sum_{\l\blQ}E_{\l\blQ}\hat{X}^{\dag}_{\l\blQ}\hat{X}_{\l\blQ}$. 
However, the eigenvalues $E_{\l\blQ}$ depend on the irreducible HSEX 
kernel which, in turn, is the difference between screened and 
bare scattering amplitudes, see Eq.~(\ref{hsex}). 
As we see below, avoiding the double counting of the (already 
incorporated) screening 
 in a perturbative theory of bosonized excitons and phonons 
is  a complex and delicate task. We therefore stick to the 
Green's function many-body theory, from which Eq.~(\ref{eigeneqX}) 
naturally emerges when solving the Bethe-Salpeter equation in the 
HSEX approximation~\cite{strinati_application_1988,onida_electronic_2002}.

The time-dependent average of $\hat{N}_{\l \blQ}$ can be calculated from the 
2-particle Green's function~\cite{svl-book}
\begin{align}
G_{2}(iz_{i},jz_{j};mz_{m},nz_{n})\equiv &-
\bra\callT\big\{
\hat{d}_{i}(z_{i})\hat{d}_{j}(z_{j})\hat{d}^{\dag}_{n}(z_{n})
\hat{d}^{\dag}_{m}(z_{m})\big\}\ket,
\label{G2}
\end{align}
where the $z$'s are times on the Keldysh contour 
 $C=C_{-}\cup C_{+}=(0,\iif)\cup (\iif,0)$ and $\callT$ is the 
contour ordering operator. The indices $i,j,m,n$ carried by the 
electronic 
operators in Eq.~(\ref{G2}) specify 
band and momentum.
Consider the two-time propagator
\begin{align}
N^{\blQ}_{cv\blk,c'v'\blk'}(z,z')&\equiv -
G_{2}(c\blk+\blQ z,v'\blk' z';v\blk z^{+},c'\blk'+\blQ z'^{+})
\nn\\
&=\bra \callT\big\{ 
\hat{d}^{\dag}_{v\blk}(z)\hat{d}_{c\blk+\blQ}(z)
\hat{d}^{\dag}_{c'\blk'+\blQ}(z')\hat{d}_{v'\blk'}(z')\big\}\ket.
\label{NQcv}
\end{align}
Rotating this quantity in the excitonic basis we obtain the {\em exciton 
Green's function}
\begin{align}
N_{\l\blQ}(z,z')&=\sum_{\substack{cv\blk\\c'v'\blk'}}
A^{\l \blQ\ast}_{cv\blk}\,N^{\blQ}_{cv\blk,c'v'\blk'}(z,z')
A^{\l \blQ}_{c'v'\blk'}
=\bra\callT\big\{\hat{X}_{\l\blQ}(z)\hat{X}^{\dag}_{\l'\blQ}(z')\big\}\ket.
\label{Nllp}
\end{align}
For any real time $t$ we use the notation  
$z=t_{\pm}$ if $z\in C_{\pm}$.
The average number of excitons is simply given by
\begin{align}
N_{\l\blQ}(t)=N_{\l\blQ}(t_{-},t_{+})=N^{<}_{\l\blQ}(t,t).
\label{xn}
\end{align}

It is useful to write the two-particle Green's function according 
to
\begin{align}
G_{2}(iz_{i},jz_{j};mz_{m},nz_{n})&=
G_{im}(z_{i},z_{m})G_{jn}(z_{j},z_{n})
-L(iz_{i},jz_{j};mz_{m},nz_{n}),
\label{g2=gg-l}
\end{align}
where 
$G_{im}(z_{i},z_{m})=-i\bra\callT\{\hat{d}_{i}(z_{i})\hat{d}^{\dag}_{m}(z_{m})\big\}\ket$ 
is the one-particle Green's function and 
$L$ is the so called exchange-correlation (xc) 
function~\cite{svl-book}, which can be determined diagrammatically 
from the solution of the 
Bethe-Salpeter equation. Inserting 
this expression in Eq.~(\ref{NQcv}) we obtain
\begin{align}
N^{\blQ}_{cv\blk,c'v'\blk'}(z,z')=\d_{\blQ,\bz}\r_{cv\blk}(t)
\r_{v'c'\blk'}(t')+ N^{{\rm inc},\blQ}_{cv\blk,c'v'\blk'}(z,z'),
\label{N=c+i}
\end{align}
where 
\begin{align}
\r_{\m\n\blk}(t)\equiv-i
G_{\m\n\blk}(z, z^{+})=
\bra \hat{d}^{\dag}_{\n\blk}(z)\hat{d}_{\m\blk}(z)\ket
\end{align}
is the one-particle density matrix, and hence
\begin{align}
N^{{\rm inc},\blQ}_{cv\blk,c'v'\blk'}(z,z')=
L^{\blQ}_{cv\blk,c'v'\blk'}(z,z';z^{+},z'^{+})\equiv
L(c\blk+\blQ z,v'\blk' z';v\blk z^{+},c'\blk'+\blQ z'^{+}) 
.
\label{dN}
\end{align}
Inserting Eq.~(\ref{N=c+i}) into Eq.~(\ref{Nllp}), the exciton 
Green's function reads
\begin{align}
N_{\l\blQ}(z,z')=\d_{\blQ,\bz}|\r_{\l}(t)|^{2}+
N^{{\rm inc}}_{\l\blQ}(z,z'),
\label{N=c+ixb}
\end{align}
where the {\em exciton polarization} is defined according to
\begin{align}
\r_{\l}(t)\equiv \sum_{cv\blk}A^{\l 
\bz\ast}_{cv\blk}\r_{cv\blk}(t)=\bra\hat{X}_{\l \bz}(t)\ket,
\label{xpol}
\end{align}
and the incoherent exciton Green's function 
$N^{{\rm inc}}_{\l\blQ}(z,z')$ is defined as in Eq.~(\ref{Nllp}) 
with\linebreak $N\to N^{{\rm inc}}$. 
As the exciton 
wavefunction is normalized to unity, 
the exciton polarization scales like $\sqrt{\callN_{\blk}}$, as it 
should be.
Equation~(\ref{N=c+ixb}) shows that the number of 
excitons, see Eq.~(\ref{xn}), is naturally written as the sum of the number of {\em 
coherent excitons} $\d_{\blQ,\bz}|\r_{\l}(t)|^{2}$ and the number of {\em 
incoherent excitons} $N^{{\rm inc}}_{\l\blQ}(t)\equiv N^{{\rm 
inc},<}_{\l\blQ}(t,t)$, i.e.,
\begin{align}
N_{\l\blQ}(t)=
\d_{\blQ,\bz}|\r_{\l}(t)|^{2}+
N^{{\rm inc}}_{\l\blQ}(t).
\label{xdm}
\end{align}
Coherent excitons are generated 
by optical excitations and therefore have zero momentum.

\subsection{Ab initio Hamiltonian}
\label{aihamsec}

The dynamics of any quasiparticle in a crystal, including the 
exciton, is governed by the first principles 
Hamiltonian for electrons and phonons
\begin{align}
\hat{H}(t)=\hat{H}_{\rm crystal}+\hat{H}_{\rm drive}(t).
\end{align}
The unperturbed crystal Hamiltonian is written in terms
of normal mode displacement 
$\hat{U}_{\a\blQ }$ and momentum $\hat{P}_{\a\blQ }$ operators of the 
nuclear lattice, satisfying the commutation relations $[\hat{U}_{\a\blQ},
\hat{P}_{\a'\blQ' }^{\dag}]=\d_{\blQ,\blQ'}\d_{\a\a'}$, as well as 
electronic field operators $\hat{d}_{\m\blk}$. We have~\cite{stefanucci_in-and-out_2023}
\begin{align}
\hat{H}_{\rm crystal}=\hat{H}_{0,e}+\hat{H}_{0,ph}+
\hat{H}_{e-e}+\hat{H}_{e-ph},
\label{el-phonham3}
\end{align}
where
\begin{subequations}
\begin{align}
\hat{H}_{0,e}&=\sum_{\blk\m\m'}h_{\m\m'}(\blk)\hat{d}^{\dag}_{\m\blk}\hat{d}_{\m'\blk},
\label{h0es}
\\
\hat{H}_{0,ph}&=\frac{1}{2}\sum_{\blQ\a\a'}
\big(\hat{U}_{\a\blQ}^{\dag},\hat{P}_{\a\blQ}^{\dag}\big)
\left(\!\!\begin{array}{cc}
	\k_{\a\a'}(\blQ) & 0 \\ 0 & \d_{\a\a'} 
\end{array}\!\!\right)
\left(\begin{array}{cc}
	\hat{U}_{\a'\blQ} \\ \hat{P}_{\a'\blQ} 
\end{array}\right)
-\sum_{\blk\m\m'}\sum_{\a}
\r^{\rm eq}_{\m'\m\blk}
g^{\m\m'}_{\a{\mathbf 0}}(\blk)\,
\hat{U}_{\a{\mathbf 0}},
\label{h0phs}
\\
\hat{H}_{e-e}&=
\frac{1}{2}\sum_{\substack{\blk\blk'\blQ\\ \m\m'\n\n'}}
\!\! v_{\m\blk+\blQ\,\n'\blk'-\blQ\,\n\blk'\,\m'\blk}
\hat{d}^{\dag}_{\m\blk+\blQ}\hat{d}^{\dag}_{\n'\blk'-\blQ}
\hat{d}_{\n\blk'}\hat{d}_{\m'\blk},
\label{hees}
\\
\hat{H}_{e-ph}&=
\sum_{\blk\m\m'}\sum_{\blQ\a}
\hat{d}^{\dag}_{\m\blk}\hat{d}_{\m'\blk-\blQ}
g_{\a-\blQ}^{\m\m'}(\blk)\,
\hat{U}_{\a\blQ}.
\label{hephs}
\end{align}
\label{hcomp}
\end{subequations}
In Eq.~(\ref{h0es}), 
$h_{\m\m'}(\blk)=\bra\m\blk|\frac{\hat{\blp}^{2}}{2}+V(\hat{\blr})|\m'\blk\ket$ 
is the matrix element of the one-electron Hamiltonian, $V(\blr)$ 
being the potential generated by the nuclei in their equilibrium 
positions. Equation~(\ref{h0phs}) is the Hamiltonian of the bare 
phonons, with $\k_{\a\a'}(\blQ)$ the elastic tensor, \linebreak 
$\r^{\rm eq}_{\m'\m\blk}=\bra 
\hat{d}^{\dag}_{\m\blk}\hat{d}_{\m'\blk}\ket$ the equilibrium 
density matrix, and 
\begin{align}
g_{\a-\blQ}^{\m\m'}(\blk)=\bra\m\blk|\left.\frac{\de V(\hat{\blr})}
{\de U_{\a\blQ}}\right|_{\blU_{\a\blQ}=0}|\m'\blk-\blQ\ket=
g^{\m'\m\ast}_{\a\blQ}(\blk-\blQ)
\label{ephcoupprop}
\end{align}
the {\em bare} $e$-$ph$ couplings. 
Notice that Eq.~(\ref{ephcoupprop}) 
scales like $1/\sqrt{\callN_{\blk}}$ since the commutation relation between 
$\hat{U}_{\a\blQ}$ and $\hat{P}^{\dag}_{\a'\blQ'}$ has been normalized to a 
Kronecker delta.
The second 
term in Eq.~(\ref{h0phs}) guarantees 
that the time-derivatives of the 
nuclear momenta vanish in equilibrium~\cite{stefanucci_in-and-out_2023}.
The elastic tensor satisfies the exact identity 
(in the basis of the Born-Oppenheimer normal 
modes)~\cite{feliciano_electron-phonon_2017,baroni_phonons_2001,stefanucci_in-and-out_2023}
\begin{align}
\k_{\a\a'}(\blQ)+\P^{\rm ph}_{\blQ\a\a'}(\w=0)=\d_{\a\a'}\w^{2}_{\a\blQ},
\label{exactid}
\end{align}
where $\P^{\rm ph}_{\a\blQ}(\w)$ is the equilibrium phononic self-energy in the 
clamped-nuclei approximation, and 
$\w^{2}_{\a\blQ}$ are the eigenvalues of the Hessian of the Born-Oppenheimer 
energy. The $e$-$e$ and $e$-$ph$ interactions are 
described by Eqs.~(\ref{hees}) and (\ref{hephs}) respectively.
The full ab initio Hamiltonian can alternatively be written in 
terms of the Born-Oppenheimer phononic operators $\hat{b}_{\a\blQ}$ using the 
relations
\begin{align}
\hat{U}_{\a\blQ}=\frac{1}
{\sqrt{2\w_{\a\blQ}}}\,(\hat{b}_{\a\blQ}+\hat{b}^{\dag}_{\a-\blQ})\quad,\quad
\hat{P}_{\a\blQ}=-i\sqrt{\frac{\w_{\a\blQ}}{2}}\,
(\hat{b}_{\a\blQ}-\hat{b}^{\dag}_{\a-\blQ}).
\end{align}

The driving Hamiltonian accounts for the light-matter interaction and 
reads
\begin{align}
\hat{H}_{\rm 
drive}=\sum_{\m\n\blk}\W_{\m\n\blk}(t)\hat{d}^{\dag}_{\m\blk}\hat{d}_{\n\blk},
\end{align}
where
\begin{align}
\W_{\m\n\blk}(t)\equiv \frac{1}{2c}\bra\m\blk|\hat{\blp}\cdot 
\blA(\hat{\blr},t)+\blA(\hat{\blr},t)\cdot\hat{\blp}
+\frac{1}{c}\blA^{2}(\hat{\blr},t)|\n\blk\ket
\end{align}
can be thought of as a time-dependent Rabi frequency, $\blA$ being 
the vector potential.

\subsection{The exchange-correlation function}

In this section we introduce the basic mathematical tools for the 
development of the theory.
The {\em exact} xc function $L^{\blQ}_{\m\n\blk,\m'\n'\blk'}$, see 
Eq.~(\ref{g2=gg-l}), satisfies the Bethe-Salpeter equation~\cite{stefanucci_in-and-out_2023} 
\begin{align}
L=\widetilde{L}-i \widetilde{L}(v+gD_{0}g)L,
\label{DysonL}
\end{align}
which can be represented diagrammatically as 
\begin{align}
&\raisebox{-6pt}{\includegraphics[width=0.95\textwidth]{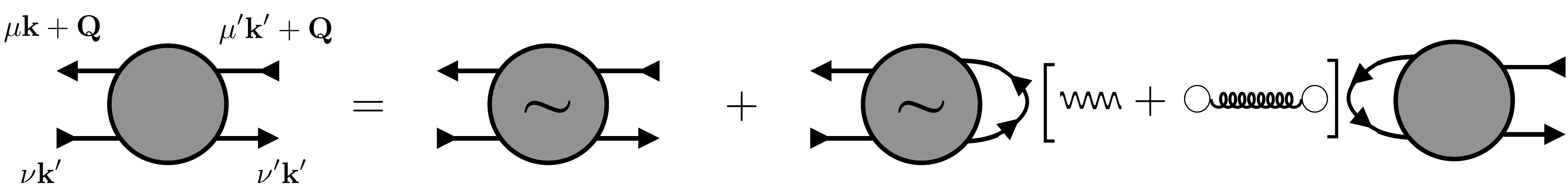}} 
\;\;.
\label{xcL}
\end{align}	
In Eq.~(\ref{DysonL}) we have the bare $e$-$e$ coupling $v$ (wiggly 
line), the bare 
$e$-$ph$ coupling $g$ (circle) and the noninteracting phonon Green's function 
$D_{0}$ (spring). We recall that $D_{0}$ does not have poles at the 
physical phonon frequencies~\cite{stefanucci_in-and-out_2023}. The latter emerge when dressing the 
phonons with electrons, see again Eq.~(\ref{exactid}). 
The correlator $\widetilde{L}$ is irreducible with respect to a cut of an 
$e$-$e$ interaction line and/or a phonon line.

Let us introduce the $v$-reducible and $D$-irreducible correlator
\begin{align}
L^{(v)}\equiv \widetilde{L}-i \widetilde{L}v L^{(v)}.
\label{DysonLcl}
\end{align}
Iterating Eq.~(\ref{DysonL}), and grouping terms with the same power 
of $g$, we get the 
{\em exact} identity
\begin{align}
L&=(\widetilde{L}-i\widetilde{L}v\widetilde{L}+\ldots)
-i(\widetilde{L}-i\widetilde{L}v\widetilde{L}+\ldots)gD_{0}g
(\widetilde{L}-i\widetilde{L}v\widetilde{L}+\ldots)
\nn\\
&-i(\widetilde{L}-i\widetilde{L}v\widetilde{L}+\ldots)gD_{0}g
(-i)(\widetilde{L}-i\widetilde{L}v\widetilde{L}+\ldots)
gD_{0}g
(\widetilde{L}-i\widetilde{L}v\widetilde{L}+\ldots)+\ldots
\nn\\
&=L^{(v)}-i\widetilde{L}g^{s}D_{0}g^{s}\widetilde{L}
-i\widetilde{L}g^{s}D_{0}g(-i)\widetilde{L}g^{s}D_{0}g^{s}\widetilde{L}+\ldots
\nn\\
&=L^{(v)}-i\widetilde{L}g^{s}\big(
D_{0}+D_{0}(-i)g\widetilde{L}g^{s}D_{0}+\ldots)g^{s}\widetilde{L}
\nn\\
&=L^{(v)}-i\widetilde{L}g^{s}Dg^{s}\widetilde{L}.
\label{Lwo}
\end{align}
In the second equality we have recognized  
the {\em screened} $e$-$ph$ coupling 
$g^{s}=(1-i vL^{(v)})g$\linebreak $=(1-iW\widetilde{L})g$, defined diagrammatically by the 
equation 
below~\cite{feliciano_electron-phonon_2017,stefanucci_in-and-out_2023}:
\begin{align}
\raisebox{-6pt}{\includegraphics[width=0.5\textwidth]{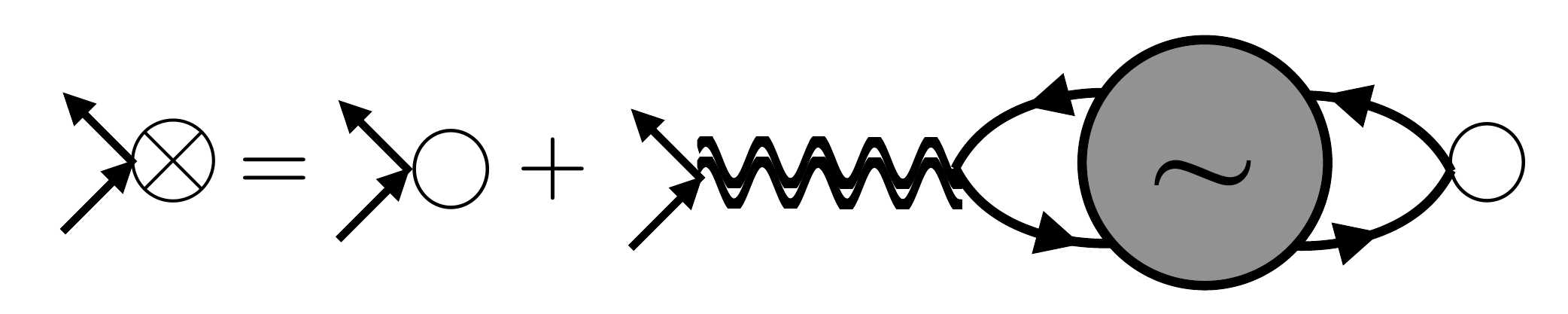}} 
\;\;,
\label{gscreened}
\end{align}
where the circle is $g$, the crossed circle is $g^{s}$ and the double 
wiggly line is the screened Coulomb interaction 
$W=v-iv\widetilde{L}W$.
In the third equality we have recognized the  phononic 
self-energy $\P^{\rm ph}=-i 
g\widetilde{L}g^{s}$~\cite{feliciano_electron-phonon_2017,stefanucci_in-and-out_2023}, 
which, in the forth equality, has been used to dress the phonon Green's function 
$D=D_{0}+D_{0}\P^{\rm ph}D$. 
Taking into account that 
$g^{s}\tilde{L}=g L^{(v)}$, the xc 
function can alternatively be written as
\begin{align}
L=L^{(v)}-iL^{(v)}gDgL^{(v)}=L^{(v)}-i\widetilde{L}g^{s}DgL^{(v)}=
L^{(v)}-iL^{(v)}gDg^{s}\widetilde{L}.
\label{altL0}
\end{align}
Henceforth we refer to the two-time functions
\begin{align}
\widetilde{N}^{\blQ}_{cv\blk,c'v'\blk'}(z,z')=
\widetilde{L}^{\blQ}_{cv\blk,c'v'\blk'}(z,z';z^{+},z'^{+}),
\label{Ntilde}
\end{align}
as  the irreducible exciton propagator, 
\begin{align}
N^{(v)\blQ}_{cv\blk,c'v'\blk'}(z,z')=
L^{(v)\blQ}_{cv\blk,c'v'\blk'}(z,z';z^{+},z'^{+}),
\label{Ntilde}
\end{align}
as  the $v$-reducible exciton propagator, 
and 
\begin{align}
\widetilde{N}^{(D)\blQ}_{cv\blk,c'v'\blk'}(z,z')=-i
[\widetilde{L}g^{s}Dg^{s}\widetilde{L}]^{\blQ}_{cv\blk,c'v'\blk'}(z,z';z^{+},z'^{+}),
\label{ND}
\end{align}
as the $D$-reducible exciton 
propagator.

Excitonic effects in photoabsorption spectra are captured by the 
following approximation:
\begin{equation}
\widetilde{L}\simeq \widetilde{L}^{\rm SEX}=\ell+i \ell K^{(r),\rm SEX} \,\ell,
\label{Lirrapp}
\end{equation}
which can be represented diagrammatically as 
\begin{align}
&\raisebox{-6pt}{\includegraphics[width=0.55\textwidth]{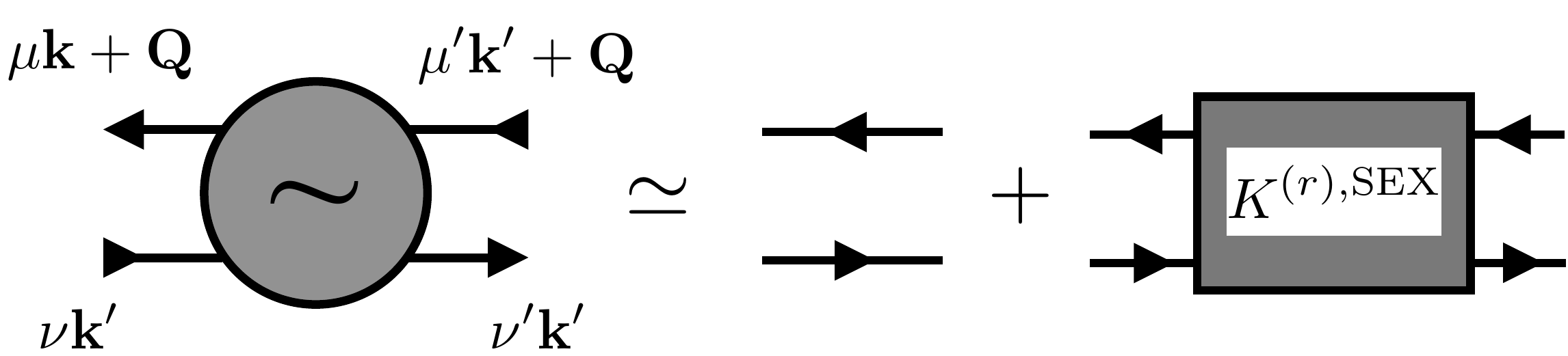}} 
\;\;.
\label{irrL}
\end{align}	
In this equation $\ell=GG$ 
is the free electron-hole propagator, whereas $K^{(r),\rm SEX}$  is 
the $\ell$-reducible 
SEX kernel. The latter solves the $T$-matrix equation
\begin{align}
K^{(r),\rm SEX}=W+i W\ell K^{(r),\rm SEX},
\label{tmatelem}
\end{align}
or, diagrammatically,
\begin{align}
&\raisebox{-6pt}{\includegraphics[width=0.55\textwidth]{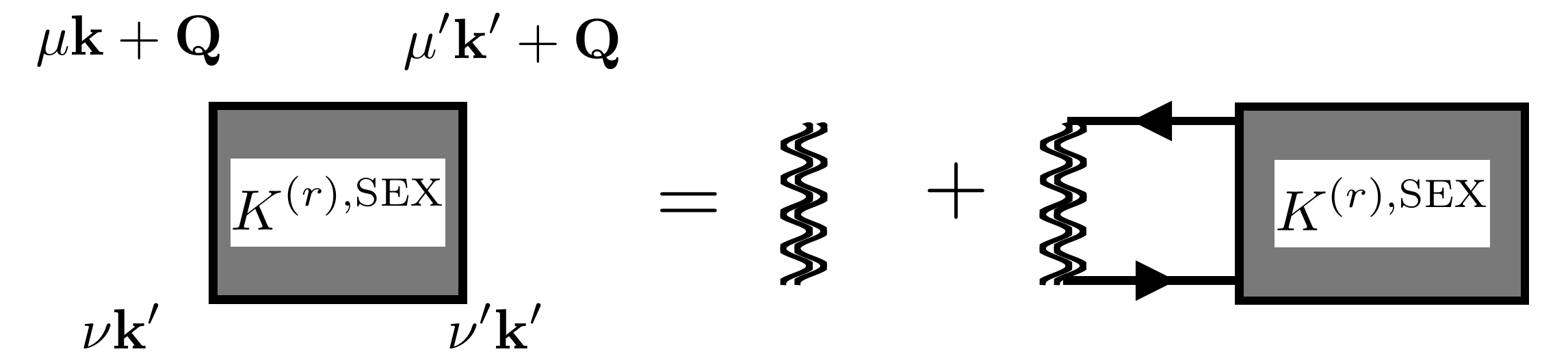}} 
\;\;,
\label{Tmat}
\end{align}	
where $W$ is the statically screened Coulomb interaction.
Inserting Eq.~(\ref{tmatelem}) into Eq.~(\ref{Lirrapp}) we can write 
$\widetilde{L}^{\rm SEX}=\ell+i \ell W \widetilde{L}^{\rm SEX}$, 
which corresponds to approximate $L^{(v)}$ in
Eq.~(\ref{DysonLcl}) as
\begin{align}
	L^{(v)}\simeq L^{\rm HSEX}=\ell+i 
	\ell K^{\rm HSEX} L^{\rm HSEX}=\ell+i \ell K^{(r),\rm 
	HSEX}\ell,
\label{elL}
\end{align}
with $K^{(r),\rm HSEX}=K^{\rm HSEX}+iK^{\rm HSEX} \ell K^{(r),\rm HSEX}$
the $\ell$-reducible 
HSEX kernel, and $K^{\rm HSEX}=W-v$ the $\ell$-irreducible 
HSEX kernel defined in Eq.~(\ref{hsex}). The diagrammatic 
representation of $K^{(r)\rm HSEX}$ is given by the equation below
\begin{align}
&\raisebox{-6pt}{\includegraphics[width=0.9\textwidth]{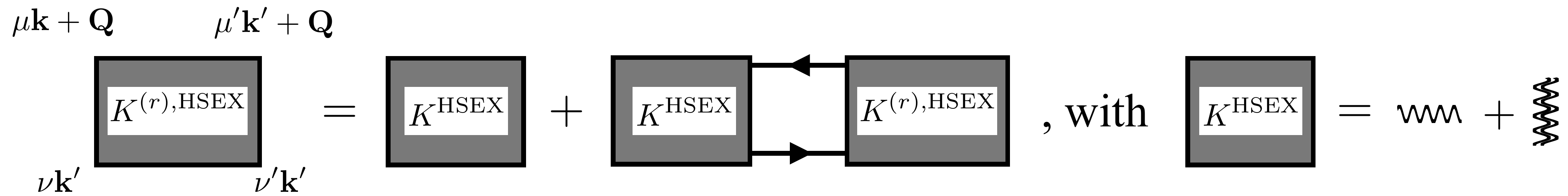}} .
\label{Kr}
\end{align}	

In $L^{\rm HSEX}$ the particle-hole 
scatters through the direct (with bare 
$v$) and exchange (with screened $W$) channels.
Solving 
Eq.~(\ref{elL}) in equilibrium is equivalent to solving 
Eq.~(\ref{eigeneqX}). In fact, the so called Bethe-Salpeter 
Hamiltonian is nothing but the ``pole'' of (the retarded) $L^{\rm HSEX}$. 
Generally speaking, elevating the pole of a quasiparticle correlator 
(in our case, the exciton) to the status of a Hamiltonian
is not recommended when the 
quasiparticles interact with other degrees of freedom (such as 
the phonons in our case).

Let us denote by $N^{\rm HSEX}$ the two-time 
propagator in Eq.~(\ref{dN}) evaluated with $L=L^{\rm HSEX}$. 
Taking into account that the free electron-hole propagator 
$\ell^{\blQ}_{cv\blk,c'v'\blk'}(z,z')=\d_{\blk\blk'}\d_{cc'}\d_{vv'}
g_{c\blk+\blQ}(z,z')$\linebreak$g_{v\blk}(z',z)$, with $g_{i\blk}$ 
the quasi-particle Green's function, the equations of motion for the 
HSEX propagator read
\begin{subequations}
\begin{align}
\Big[i\frac{d}{dz}-E^{\blQ}_{cv\blk}\Big]N^{\rm 
HSEX,\blQ}_{cv\blk,c'v'\blk'}(z,z')+(f^{\rm el}_{v\blk}-f^{\rm 
el}_{c\blk+\blQ})\sum_{c_{1}v_{1}\blk_{1}}K^{\rm HSEX,\blQ}_{cv\blk,c_{1}v_{1}\blk_{1}}
N^{\rm 
HSEX,\blQ}_{c_{1}v_{1}\blk_{1},c'v'\blk'}(z,z')
\nn\\
=i\d_{\blk\blk'}\d_{cc'}\d_{vv'}\d(z,z')
(f^{\rm el}_{v\blk}-f^{\rm 
el}_{c\blk+\blQ}),
\end{align}	
\begin{align}
\Big[-i\frac{d}{dz'}-E^{\blQ}_{c'v'\blk'}\Big]N^{\rm 
HSEX,\blQ}_{cv\blk,c'v'\blk'}(z,z')+(f^{\rm el}_{v'\blk'}-f^{\rm 
el}_{c'\blk'+\blQ})\sum_{c_{1}v_{1}\blk_{1}}
N^{\rm 
HSEX,\blQ}_{cv\blk,c_{1}v_{1}\blk_{1}}(z,z')
K^{\rm HSEX,\blQ}_{c_{1}v_{1}\blk_{1},c'v'\blk'}
\nn\\
=i\d_{\blk\blk'}\d_{cc'}\d_{vv'}\d(z,z')
(f^{\rm el}_{v\blk}-f^{\rm 
el}_{c\blk+\blQ}),
\end{align}	
\label{nhsexeom}
\end{subequations}
where $f^{\rm el}_{\m\blk}$ the electronic occupation of band $\m$ at 
momentum $\blk$. For small excitation densities, the Pauli blocking 
factors $(f^{\rm el}_{v'\blk'}-f^{\rm 
el}_{c'\blk'+\blQ})$ multiplying the kernel are approximated as unity. 
These factors are responsible for 
the renormalization 
of the exciton energies and wavefunctions. Rotating 
Eqs.~(\ref{nhsexeom}) in the excitonic basis, that is
\begin{align}
N^{\rm HSEX,\blQ}_{cv\blk,c'v'\blk'}(z,z')=\sum_{\l}A^{\l \blQ}_{cv\blk} \,
N^{\rm HSEX}_{\l\blQ}(z,z')\,
A^{\l \blQ\ast}_{c'v'\blk'}\;,
\label{N(v)}
\end{align}
we obtain
\begin{subequations}
\begin{align}
\Big[i\frac{d}{dz}-E_{\l\blQ}\Big]N^{\rm HSEX}_{\l\blQ}(z,z')
=i\d(z,z')\sum_{cv\blk}|A^{\l \blQ}_{cv\blk}|^{2}
(f^{\rm el}_{v\blk}-f^{\rm 
el}_{c\blk+\blQ}),
\end{align}	
\begin{align}
\Big[-i\frac{d}{dz'}-E_{\l\blQ}\Big]N^{\rm HSEX}_{\l\blQ}(z,z')
=i\d(z,z')\sum_{cv\blk}|A^{\l \blQ}_{cv\blk}|^{2}
(f^{\rm el}_{v\blk}-f^{\rm 
el}_{c\blk+\blQ}).
\end{align}	
\label{nhsexeom2}
\end{subequations}
Accordingly, the propagator
$N^{\rm HSEX}(z,z')$ does not 
contribute to the number of incoherent excitons 
since
\begin{align}
\frac{d}{dt}N^{\rm HSEX,<}_{\l\blQ}(t,t)=0.
\label{nhsex<=0}
\end{align}

In the following sections we focus on small excitation densities, 
considering contributions beyond the linear regime.  
Some second-order effects must be 
discarded to achieve a closed set of equations. These include the 
aforementioned renormalization 
of the exciton energies and wavefunctions due to Pauli blocking 
factors as well as the update of the screened Coulomb interaction 
$W$ during the time evolution. We also assume that the 
Tamm-Dancoff approximation for the solution of the Bethe-Salpeter 
equation remains accurate. Although these approximations are 
ubiquitous in the literature, it would be interesting to 
investigate 
their impact  in the future.

\section{Coherent excitons}
\label{cohxsec}

The equation of motion for the exciton polarization 
$\r_{\l}(t)$ follows from the equation of motion of 
$\r_{cv\blk}(t)$~\cite{stefanucci_semiconductor_2024}:
\begin{align}
\frac{d}{dt}\r_{cv\blk}&+i\big(\e_{c\blk}-\e_{v\blk}-i\G^{\rm pol}_{cv\blk}\big)\r_{cv\blk}
-i\sum_{c'v'\blk'}
K^{\rm HSEX,\bz}_{cv\blk,c'v'\blk'}\r_{c'v'\blk'}
=-i\W_{cv\blk}.
\label{sbepol}
\end{align}
This equation reduces to the time-dependent HSEX equation for $\G^{\rm 
pol}_{cv\blk}=0$~\cite{perfetto_pump-driven_2019,sangalli_excitons_2021,chan_giant_2023}. In the linear response regime, the time-dependent HSEX equations
are equivalent to solving the Bethe-Salpeter 
equation~\cite{attaccalite_real-time_2011,svl-book,perfetto_nonequilibrium_2015}. 
In fact, the exciton polarization is proportional to the induced 
field, and therefore its Fourier transform is connected to the 
absorption spectrum. 

The polarization rates $\G^{\rm pol}_{cv\blk}$ 
 can be calculated by different 
means~\cite{toyozawa_interband_1964,marini_ab-initio_2008,chan_exciton_2023}, 
although they
are often treated as fitting 
parameters. 
For small excitation densities  $\G^{\rm pol}_{cv\blk}$  is
dominated by $e$-$ph$ scattering 
mechanisms~\cite{perfetto_real_2022,perfetto_real-time_2023}.
The polarization rates in the Fan-Migdal approximation  can be obtained 
using the 
mirrored form of the Generalized 
Kadanoff-Baym Ansatz 
(MGKBA)~\cite{stefanucci_semiconductor_2024,kalvova_dynamical_2023} 
(the MGKBA corrects the standard GKBA which leads to 
unphysical polarization rates).

\begin{figure}[t]
    \centering
\includegraphics[width=0.98\textwidth]{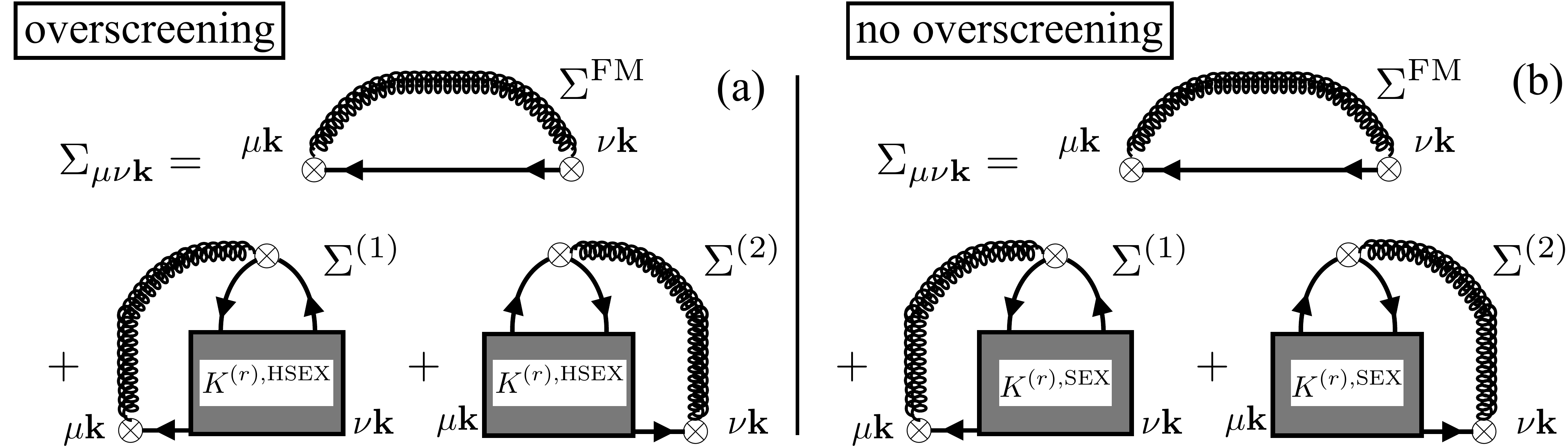}
\caption{Electronic self-energy with excitonic effects in terms of 
electronic Green's functions $G$ (solid lines), phononic 
Green's functions $D$ (double springs), and screened $e$-$ph$ 
couplings $g^{s}$
(circled crosses). Panel (a) shows $\S$ affected by overscreening.  
Panel (b) shows $\S$ with no overscreening.}
\label{Sigma-el}
\end{figure}

The treatment of Ref.~\cite{stefanucci_semiconductor_2024} must be 
improved for semiconductors hosting excitons. The origin of the term 
$\G^{\rm pol}_{cv\blk}\r_{cv\blk}$ stems from the collision integral
\begin{align}
\G^{\rm pol}_{cv\blk}(t)\r_{cv\blk}(t)=S_{cv\blk}(t)=
\int d\bar{z}\big[\S(z,\bar{z})G(\bar{z},z^{+})-
G(z,\bar{z})\S(\bar{z},z^{+})\big]_{cv\blk},
\label{xphscatt}
\end{align}
where $\S$ is the electronic correlation energy. 
In the presence of excitons, the Fan-Migdal self-energy alone (see 
the first diagram in Fig.~\ref{Sigma-el}) is not sufficient  
because electrons and holes cannot form bound states.     
A suitable approximation for $\S$ can be deduced from those 
approaches that treat excitons as composite bosonic particles~\cite{chan_exciton_2023}. In 
these approaches 
the {\em coherent-exciton} self-energy has the structure  $\P^{\rm 
ex}(z,z')=i\callG 
D(z,z') N^{\rm HSEX}(z,z') \callG$, where 
the  
{\em exciton-phonon 
coupling}~\cite{toyozawa_theory_1958,thranhardt_quantum_2000,jiang_exciton-phonon_2007,chen_exciton-phonon_2020,cudazzo_first-principles_2020,antonius_theory_2022}
\begin{align}
\callG^{\l\l'}_{\a-\blQ'}(\blQ)\equiv 
\sum_{c_{1}c_{2}v_{1}\blk_{1}}
A^{\l\blQ\ast}_{c_{1}v_{1}\blk_{1}}g^{s,c_{1}c_{2}}_{\a-\blQ'}(\blk_{1}+\blQ)
A^{\l'\blQ-\blQ'}_{c_{2}v_{1}\blk_{1}}
-
\sum_{c_{1}v_{1}v_{2}\blk_{1}}A^{\l\blQ\ast}_{c_{1}v_{1}\blk_{1}}
g^{s,v_{2}v_{1}}_{\a-\blQ'}(\blk_{1}+\blQ')A^{\l'\blQ-\blQ'}_{c_{1}v_{2}\blk_{1}+\blQ'}
\label{callGx2}
\end{align}
depends on the {\em screened} $e$-$ph$ coupling $g^{s}$ 
and the exciton wavefunctions defined in Eq.~(\ref{eigeneqX}).

In the ab initio formulation, 
the polarization rates generated by the model $\P^{\rm ex}$ are 
produced by the {\em electronic} self-energy in 
Fig.~\ref{Sigma-el}(a), see below for the proof of this statement.
However, the screened $e$-$ph$ coupling
in  
Fig.~\ref{Sigma-el}(a)  leads to an overscreening. Consider, for 
instance, the second diagram $\S^{(1)}$. Taking into account that 
$W=v-i v\widetilde{L}^{\rm SEX}W$,
and using the explicit form of $g^{s}$ for the $e$-$ph$ coupling 
at the top of the 
diagram (with $\widetilde{L}\simeq \widetilde{L}^{\rm SEX}$), we get the 
 diagrammatic structure
\begin{align}
g^{s}\ell K^{(r),\rm HSEX}&=g\ell K^{(r),\rm HSEX}-i
g\widetilde{L}^{\rm SEX}W\ell K^{(r),\rm HSEX}
\nn\\
&=g\ell K^{(r),\rm HSEX}-i g\widetilde{L}^{\rm SEX}
(v-i v\widetilde{L}^{\rm SEX}v +\ldots)
\ell (K^{\rm HSEX}+i K^{\rm HSEX}\ell K^{\rm HSEX}+\ldots)
\nn\\
&=g\ell K^{(r),\rm HSEX}-ig\widetilde{L}^{\rm SEX}(v-i v\widetilde{L}^{\rm SEX}v 
+\ldots)L^{\rm HSEX} K^{\rm HSEX}
\nn\\
&=g\ell K^{(r),\rm HSEX}-ig\widetilde{L}^{\rm SEX}(v-i v\widetilde{L}^{\rm SEX}v +\ldots)
(\widetilde{L}^{\rm SEX}-i\widetilde{L}^{\rm SEX}v\widetilde{L}^{\rm SEX}+\ldots)K^{\rm HSEX},
\label{oscalc}
\end{align}
from which the overscreening of the $e$-$ph$ coupling is evident.

To solve the overscreening problem
and simultaneously develop a theory in terms of 
 screened $e$-$ph$ couplings, it is sufficient to examine the exact 
form of the electronic
self-energy~\cite{feliciano_electron-phonon_2017,stefanucci_in-and-out_2023}. 
The closest 
approximation to Fig.~\ref{Sigma-el}(a) which is free of overscreening 
is the one in Fig.~\ref{Sigma-el}(b), where 
$K^{(r),\rm HSEX}\to K^{(r),\rm SEX}$. The crucial difference between 
the self-energies in Fig.~\ref{Sigma-el} is that 
$\S G \propto g^{s}Dg^{s}L^{\rm HSEX}$ for panel (a), see Eq.~(\ref{elL}), 
while
$\S G \propto g^{s}Dg^{s}\widetilde{L}^{\rm SEX}=g^{s}DgL^{\rm HSEX}$ for panel (b), see 
Eq.~(\ref{Lirrapp}).
To evaluate the collision integral, see Eq.~(\ref{xphscatt}), with 
the self-energy of panel (b) we
make the following ansatz for 
the off-diagonal elements of the electronic Green's 
function~\cite{ShaferWegenerbook}
\begin{subequations}
\begin{align}
G_{cv\blk}(z,z')&=-G_{cc\blk}(z,z')\r_{cv\blk}(t')+
\r_{cv\blk}(t)G_{vv\blk}(z,z'),
\\
G_{vc\blk}(z,z')&=-\r_{vc\blk}(t)G_{cc\blk}(z,z')+
G_{vv\blk}(z,z')\r_{vc\blk}(t').
\end{align}
\label{vcansatz}
\end{subequations}
This ansatz is exact at the mean-field level.
Equation~(\ref{vcansatz}) allows for expressing all matrix elements of 
the irreducible xc function $\widetilde{L}^{\rm SEX}$ 
in terms of the irreducible exciton propagator [compare 
with Eq.~(\ref{dN})] 
\begin{align}
\widetilde{N}^{\rm SEX,\blQ}_{cv\blk,c'v'\blk'}(z,z')= 
\widetilde{L}^{\rm SEX,\blQ}_{cv\blk,c'v'\blk'}
(z,z';z^{+},z'^{+}),
\label{Nirr}
\end{align}
see Appendix~\ref{polrates} for details. 
The irreducible exciton propagator takes a particularly simple 
form in the irreducible excitonic basis. Let us consider the 
eigenvalue equation [compare with Eq.~(\ref{eigeneqX})]
\begin{align}
E^{\blQ}_{cv\blk}\widetilde{A}^{\widetilde{\l}\blQ}_{cv\blk}-
\sum_{c'v'\blk'}
W^{\blQ}_{cv\blk,c'v'\blk'}\widetilde{A}^{\widetilde{\l}\blQ}_{c'v'\blk'}=
\widetilde{E}_{\widetilde{\l}\blQ}\widetilde{A}^{\widetilde{\l}\blQ}_{cv\blk}.
\label{eigeneqXirr}
\end{align}
The irreducible exciton wavefunctions $\widetilde{A}$ are chosen orthonormal for 
every $\blQ$.
Proceeding along the same lines leading to Eq.~(\ref{N(v)}) we obtain
\begin{align}
\widetilde{N}^{\rm SEX,\blQ}_{cv\blk,c'v'\blk'}(z,z')=\sum_{\widetilde{\l}}
\widetilde{A}^{\widetilde{\l} \blQ}_{cv\blk} \,
\widetilde{N}^{\rm SEX}_{\widetilde{\l}\blQ}(z,z')\,
\widetilde{A}^{\widetilde{\l} \blQ\ast}_{c'v'\blk'}.
\label{Nirrdiag}
\end{align}
The equations of motion for $\widetilde{N}^{\rm 
SEX}_{\widetilde{\l}\blQ}(z,z')$ are identical to 
Eqs.~(\ref{nhsexeom2}) with $E_{\l\blQ}\to 
\widetilde{E}_{\widetilde{\l}\blQ}$ and 
$A^{\l\blQ}\to \widetilde{A}^{\widetilde{\l}\blQ}$.
In particular, for systems in equilibrium,
\begin{align}
\widetilde{N}^{\rm SEX}_{\widetilde{\l}\blQ}(z,z')
=\th(z,z')e^{-i \widetilde{E}_{\widetilde{\l}\blQ}(t-t')}.
\label{tildeN}
\end{align}

These results can be used to rewrite the collision integral in the excitonic 
basis according to
\begin{align}
S_{\l}(z)\equiv \sum_{cv\blk}A^{\l\bz\ast}_{cv\blk}S_{cv\blk}(z)
=i\sum_{\l_{1}\l_{2}}\sum_{\a\blQ}\int\!\! d\bar{z}\;
\widetilde{\callG}^{\widetilde{\l}_{1}\l\ast}_{\a}(\blQ)
\widetilde{N}^{\rm SEX}_{\widetilde{\l}_{1}\blQ}(z,\bar{z})
\widetilde{\callG}^{\widetilde{\l}_{1}\l_{2}}_{\a}(\blQ)
\r_{\l_{2}}(\bar{t})D_{\a-\blQ}(z,\bar{z}),
\label{slambda}
\end{align}
where we define the {\em irreducible exciton-phonon 
coupling}
\begin{align}
\widetilde{\callG}^{\widetilde{\l}\l'}_{\a}(\blQ)\equiv 
\sum_{c_{1}c_{2}v_{1}\blk_{1}}
\widetilde{A}^{\widetilde{\l}\blQ\ast}_{c_{1}v_{1}\blk_{1}}g^{s,c_{1}c_{2}}_{\a-\blQ}(\blk_{1}+\blQ)
A^{\l'\bz}_{c_{2}v_{1}\blk_{1}}
-
\sum_{c_{1}v_{1}v_{2}\blk_{1}}\widetilde{A}^{\widetilde{\l}\blQ\ast}_{c_{1}v_{1}\blk_{1}}
g^{s,v_{2}v_{1}}_{\a-\blQ}(\blk_{1}+\blQ)A^{\l'\bz}_{c_{1}v_{2}\blk_{1}+\blQ}.
\label{callGx2irr}
\end{align}
Equation~(\ref{callGx2irr}) does not reduce to Eq.~(\ref{callGx2}) 
with $\blQ'=\blQ$ 
as the $e$-$ph$ coupling is contracted with the product of an exciton 
wavefunction $A$ and an irreducible exciton wavefunction 
$\widetilde{A}$. 
Assuming that the dominant contribution in Eq.~(\ref{slambda}) comes 
from the terms with $\l_{2}=\l$, we show 
in Appendix~\ref{polrates} that the Markov approximation enables us to express 
Eq.~(\ref{sbepol}) in the excitonic basis as follows 
\begin{align}
\frac{d}{dt}\r_{\l}(t)=-iE_{\l \bz}\r_{\l}(t)
-i\W_{\l}(t)- \frac{1}{2}\sum_{\widetilde{\l}'\blQ}
\widetilde{\G}^{{\rm pol}}_{\l\widetilde{\l}'\blQ}
\r_{\l}(t),
\label{drldt}
\end{align}
with $\W_{\l}\equiv \sum_{cvk}A^{\l\bz\ast}_{cv\blk}\W_{cv\blk}$. The 
polarization rates depend on  the
phononic populations \linebreak $f^{\rm ph}_{\a\blQ}\equiv \bra 
\hat{b}^{\dag}_{\a\blQ}\hat{b}_{\a\blQ}\ket$ according to
\begin{align}
\widetilde{\G}^{\rm pol}_{\l\widetilde{\l}'\blQ}&=
2\p\sum_{\a}\frac{|\widetilde{\callG}^{\widetilde{\l}'\l}_{\a}(\blQ)|^{2}}{2\w_{\a\blQ}}
\Big[\d(\widetilde{E}_{\widetilde{\l}'\blQ}-E_{\l\bz}+\w_{\a\blQ})(1+f^{\rm ph}_{\a-\blQ})+
\d(\widetilde{E}_{\widetilde{\l}'\blQ}-E_{\l\bz}-\w_{\a\blQ})f^{\rm 
ph}_{\a\blQ}\Big].
\label{gamma}
\end{align}
We  remark that $\widetilde{\G}^{\rm pol}$  involves the 
difference between an exciton energy $E$ and an irreducible exciton 
energy $\widetilde{E}$.
The necessity of introducing irreducible excitons in a theory of 
excitons and phonons has also been recognized by Paleari and 
Marini~\cite{paleari_exciton-phonon_2022}. They investigated a 
scattering channel different from the one considered here. 
Nevertheless, their study also led to the emergence of the 
irreducible exciton-phonon coupling  in 
Eq.~(\ref{callGx2irr}).       

The equation of motion Eq.~(\ref{drldt})
for the exciton polarization agrees with 
cluster expansion results~\cite{thranhardt_quantum_2000,selig_exciton_2016,brem_exciton_2018}
provided that we replace  $\widetilde{\callG}$ with  $\callG$ and 
$\widetilde{E}_{\widetilde{\l}'\blQ}$ with $E_{\l'\blQ}$. Such replacement is equivalent to 
evaluate the electronic self-energy with the diagrams of 
Fig.~\ref{Sigma-el}(a) instead of Fig.~\ref{Sigma-el}(b).
As previously discussed, this 
introduces an overscreening of the $e$-$ph$ coupling. The  
overscreening issue is common to all methods that do not rely on the 
first principles  Hamiltonian~\cite{stefanucci_in-and-out_2023}, 
but rather on model Hamiltonians where both the $e$-$e$ and 
$e$-$ph$ interactions are already 
screened. Our treatment highlights the advantages of the
first-principles Green's function formulation, where screening naturally 
emerges from the diagrammatic expansion and is therefore counted only once.
Another important feature of Eq.~(\ref{drldt}) is that it  reduces to 
the equation of motion in 
Ref.~\cite{stefanucci_semiconductor_2024} if we neglect the last two 
diagrams in Fig.~\ref{Sigma-el}.  

As a technical note, we highlight that there exists a generalization 
of Eq.~(\ref{drldt}) which agrees with more 
sophisticated treatments of the bosonic Hamiltonian for excitons and 
phonons~\cite{chan_exciton_2023}, see Appendix~\ref{polrates}.
The main 
difference with these models  is  again the occurrence of 
$\widetilde{\callG}$ and 
$\widetilde{E}$.
By retaining all terms with $\l_{2}\neq \l$ in Eq.~(\ref{slambda}) 
we show that 
$S^{\l}=\sum_{\widetilde{\l}_{1}\l_{2}\blQ}\widetilde{\G}^{\rm 
pol}_{\l\l_{2}\widetilde{\l}_{1}\blQ}\r_{\l_{2}}$ where 
$\widetilde{\G}^{\rm pol}_{\l\l_{2}\widetilde{\l}_{1}\blQ}$ 
can be obtained from Eq.~(\ref{gamma}) by 
replacing  $|\widetilde{\callG}^{\widetilde{\l}_{1}\l}_{\a}(\blQ)|^{2}\to
\widetilde{\callG}^{\widetilde{\l}_{1}\l\ast}_{\a}(\blQ)
\widetilde{\callG}^{\widetilde{\l}_{1}\l_{2}\ast}_{\a}(\blQ)$ and 
$E_{\l\bz}\to E_{\l_{2}\bz}$. It should be stressed, however, that the 
matrix $\sum_{\widetilde{\l}_{1}\blQ}\widetilde{\G}^{\rm 
pol}_{\l\l_{2}\widetilde{\l}_{1}\blQ}$ is not guaranteed to be 
positive definite.

\section{Incoherent excitons}
\label{incexsec}

\begin{figure}[t]
    \centering
\includegraphics[width=0.8\textwidth]{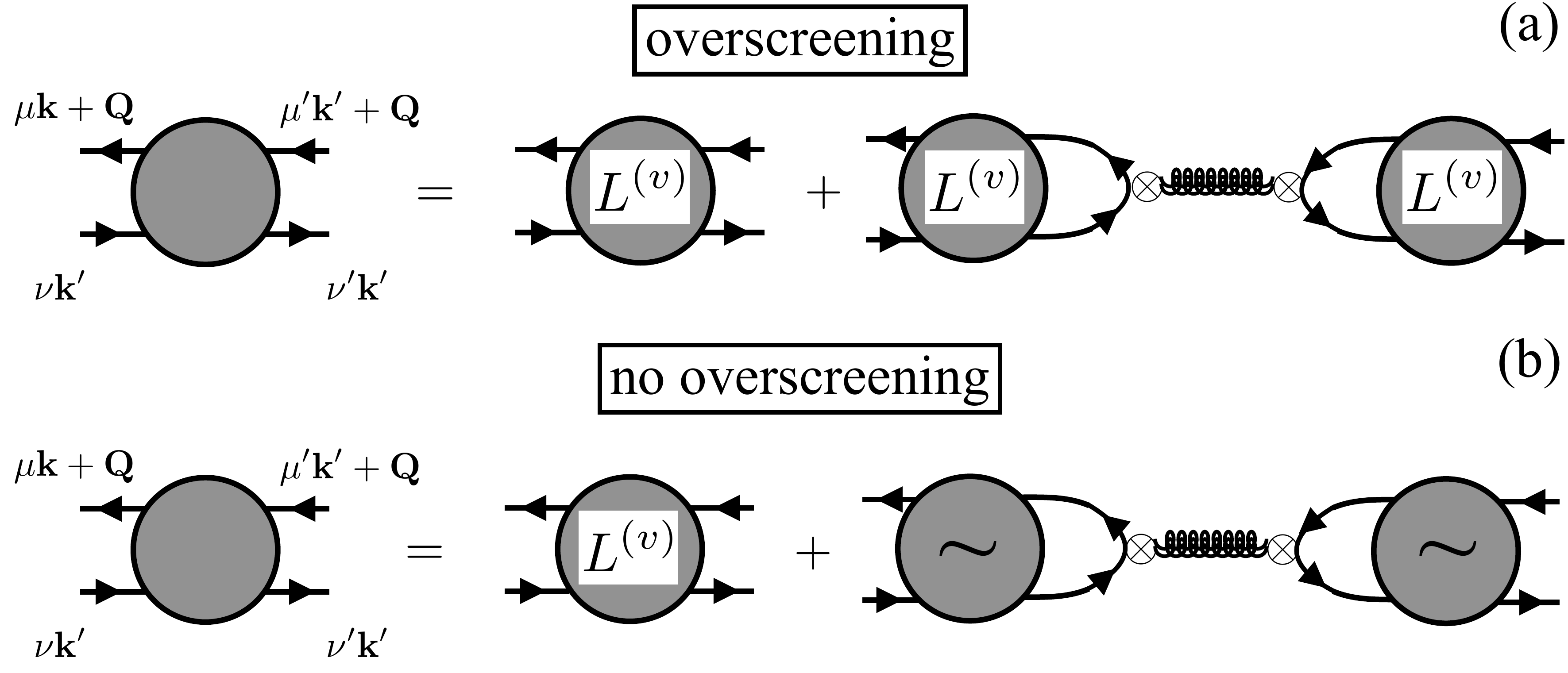}
\caption{Exchange correlation function $L$ underlying the equation of 
motion of the number of incoherent excitons. In panel (a) we show $L$ from 
model Hamiltonians whereas in panel (b) we show the exact result 
from the ab initio Hamiltonian. }
\label{Ninc}
\end{figure}

In this section we show how optically generated coherent excitons are 
converted into incoherent excitons via the $e$-$ph$ scattering.
The cluster expansion technique applied to model Hamiltonians with 
already screened $e$-$e$ and $e$-$ph$ couplings has been used 
to derive the equation of motion for incoherent excitons~\cite{thranhardt_quantum_2000,selig_exciton_2016,brem_exciton_2018}.
This equation can alternatively be derived using 
many-body Green's function methods. It is sufficient to approximate the xc function $L$ 
as in Fig.~\ref{Ninc}(a), see below. Using the same arguments as in 
the previous section, it is straightforward to realize that $L$ in 
Fig.~\ref{Ninc}(a) suffers from overscreening.   
To highlight the shortcomings of Fig.~\ref{Ninc}(a), we show in Fig.~\ref{Ninc}(b) the exact 
result from Eq.~(\ref{Lwo}).

We approximate again $\widetilde{L}\simeq \widetilde{L}^{\rm SEX}$ 
and hence $L^{(v)}\simeq L^{\rm HSEX}$.
The propagator
$N^{\rm HSEX}(z,z')$\linebreak$=L^{\rm HSEX}(z,z';z,z')$ does not 
contribute to the number of incoherent excitons, see 
Eq.~(\ref{nhsex<=0})
and discussion below it. 
We then focus on the second term of Eq.~(\ref{Lwo}), 
i.e., the $D$-reducible exciton propagator
\begin{align}
N^{(D)}\simeq 	-i\widetilde{L}^{\rm SEX}g^{s}Dg^{s}\widetilde{L}^{\rm SEX}.
\label{cxix}
\end{align}
As we see later, phonons are responsible for converting 
all coherent excitons into 
$D$-reducible excitons.
We observe that $N^{(D)}$ involves ``off-diagonal'' elements of
$\widetilde{L}^{\rm SEX}$. 
Suppose that the left $e$-$ph$ coupling has conduction 
indices. Then the left part of the diagram is 
$\widetilde{L}^{\rm SEX,\blQ}_{cv\blk,c_{1}c_{2}\blk_{1}}(z,z';z, 
z')$. On the other hand, if 
the left $e$-$ph$ coupling has valence indices 
then the left part of the diagram is 
$\widetilde{L}^{\rm SEX,\blQ}_{cv\blk,v_{2}v_{1}\blk_{1}}(z,z';z, z')$.
Analogous considerations 
apply to the right part of the diagram. 
Through the ansatz in Eqs.~(\ref{vcansatz}) 
we can express all matrix elements of $\widetilde{L}^{\rm SEX}$ in terms of 
$\widetilde{N}^{\rm SEX}$, see 
Appendix~\ref{polrates}.
Expanding $\widetilde{N}^{\rm SEX}$ in the  
irreducible excitonic basis, see Eq.~(\ref{Nirrdiag}),
and $\r$ in the excitonic basis, see Eqs.~(\ref{xpol}), 
we get
\begin{align}
N^{(D)}_{\widetilde{\l}\blQ}(z,z')=
-\int d\bar{z}d\bar{z'}\sum_{\l_{1}\l_{2}\a}
&\widetilde{\callG}^{\widetilde{\l}\l_{1}}_{\a}(\blQ)
\widetilde{\callG}^{\widetilde{\l}\l_{2}\ast}_{\a}(\blQ)
\widetilde{N}^{\rm SEX}_{\widetilde{\l}\blQ}(z,\bar{z})
\r_{\l_{1}}(\bar{t})iD_{\a\blQ}(\bar{z},\bar{z}')
\r^{\ast}_{\l_{2}}(\bar{t}')\widetilde{N}^{\rm SEX}_{\widetilde{\l}\blQ}(\bar{z}',z').
\end{align}
In the following we assume that the dominant term comes from 
$\l_{1}=\l_{2}$. For small excitation densities we can evaluate 
$\widetilde{N}^{\rm SEX}_{\widetilde{\l}\blQ}$ in equilibrium, see 
Eq.~(\ref{tildeN}), and find
\begin{subequations}
\begin{align}
\Big[i\frac{d}{dz}-\widetilde{E}_{\widetilde{\l}\blQ}\Big]	
N^{(D)}_{\widetilde{\l}\blQ}(z,z')=\int d\bar{z}\,
\P^{(D)}_{\widetilde{\l}\blQ}(z,\bar{z})
\widetilde{N}^{\rm SEX}_{\widetilde{\l}\blQ}(\bar{z},z'),
\label{eomcxix1}
\\
\Big[-i\frac{d}{dz'}-\widetilde{E}_{\widetilde{\l}\blQ}\Big]	
N^{(D)}_{\widetilde{\l}\blQ}(z,z')=\int d\bar{z}\,
\widetilde{N}^{\rm SEX}_{\widetilde{\l}\blQ}(z,\bar{z})
\P^{(D)}_{\widetilde{\l}\blQ}(\bar{z},z'),
\label{eomcxix2}
\end{align}	
\label{eomcxix}
\end{subequations}
where the self-energy for the $D$-reducible excitons reads
\begin{align}
\P^{(D)}_{\widetilde{\l}\blQ}(z,z')=\sum_{\l'\a}
|\widetilde{\callG}^{\widetilde{\l}\l'}_{\a}(\blQ)|^{2}\r_{\l'}(t)
D_{\a\blQ}(z,z')\r^{\ast}_{\l'}(t').
\label{sigmacx}
\end{align}

We extract the time-derivative of the number of $D$-reducible 
excitons 
\begin{align}
N^{(D)}(t)\equiv N^{(D),<}(t,t),
\end{align}
by taking $z=t_{-}$ 
and $z'=t'_{+}$, subtracting Eq.~(\ref{eomcxix2}) from Eq.~(\ref{eomcxix1}) 
and then setting $t=t'$:
\begin{align}
\frac{d}{dt}N^{(D)}_{\widetilde{\l}\blQ}(t)=
i\int^{t}\!\!d\bar{t}\;
\Big[\P^{(D),<}_{\widetilde{\l}\blQ}(t,\bar{t})\,
\widetilde{N}^{\rm SEX,>}_{\widetilde{\l}\blQ}(\bar{t},t)
-\P^{(D),>}_{\widetilde{\l}\blQ}(t,\bar{t})\,
\widetilde{N}^{\rm SEX,<}_{\widetilde{\l}\blQ}(\bar{t},t)\Big]+{\rm 
h.c.}\;,
\label{dncxixdt}
\end{align}
where $\widetilde{N}^{\rm SEX,\gtrless}(t,t')$ can be deduced from 
Eq.~(\ref{tildeN}).
To evaluate the collision integral in 
Eq.~(\ref{dncxixdt}) we implement again the Markov 
approximation and find 
\begin{align}
\frac{d}{dt}
N^{(D)}_{\widetilde{\l}\blQ}(t)=
\sum_{\l'}\widetilde{\G}^{\rm pol}_{\l'\widetilde{\l}\blQ}\,
|\r_{\l'}(t)|^{2},
\label{dnephdt}
\end{align}
where the polarization rates are defined in 
Eqs.~(\ref{gamma}).

By definition, see Eq.~(\ref{dN}), the number of incoherent exciton 
with quantum number $\l$ and momentum $\blQ$
produced by our approximation is 
\begin{align}
N^{\rm inc}_{\l\blQ}(t)=\sum_{\widetilde{\l}'} |S^{\blQ}_{\l\widetilde{\l}'}|^{2}
N^{(D)}_{\widetilde{\l}'\blQ}(t),
\end{align}
where
\begin{align}
S^{\blQ}_{\l\widetilde{\l}'}=\sum_{cv\blk}A^{\l\blQ\ast}_{cv\blk}\widetilde{A}^{\widetilde{\l}'\blQ}_{cv\blk}
\end{align}
is the overlap matrix between excitons and irreducible excitons. 
Therefore, the equation of motion for the number of incoherent 
excitons reads
\begin{align}
\frac{d}{dt}
N^{\rm inc}_{\l\blQ}(t)=
\sum_{\l'\widetilde{\l}'}|S^{\blQ}_{\l\widetilde{\l}'}|^{2}\widetilde{\G}^{\rm pol}_{\l'\widetilde{\l}'\blQ}\,
|\r_{\l'}(t)|^{2}.
\label{dnincdt3}
\end{align}

From the equations of motion Eqs.~(\ref{drldt}) and (\ref{dnincdt3}) 
we can easily deduce the equation of motion for the total number of 
excitons, see 
Eq.~(\ref{xdm}),
\begin{align}
N=\sum_{\l\blQ}N_{\l\blQ}=\sum_{\l\blQ}\big(\d_{\blQ,\bz}|\r_{\l}|^{2}+N^{\rm 
inc}_{\l\blQ}\big).
\end{align}
Since $\sum_{\l}|S^{\blQ}_{\l\widetilde{\l}'}|^{2}=1$ for all 
$\widetilde{\l}'$ we have $\sum_{\l}N^{\rm inc}_{\l\blQ}(t)=
\sum_{\widetilde{\l}}N^{(D)}_{\widetilde{\l}\blQ}(t)$, and 
therefore
\begin{align}
\frac{d}{dt}N=2\sum_{\l}\Im\left[\W_{\l}\r_{\l}^{\ast}\right].
\label{dNtotdt}
\end{align}	
The first principles formulation confirms an important result from model 
Hamiltonians, which states that the total number of excitons remains 
constant after the optical field is applied.

\section{Inelastic exciton-phonon scattering}
\label{xxsec}

\begin{figure}[tbp]
    \centering
\includegraphics[width=0.98\textwidth]{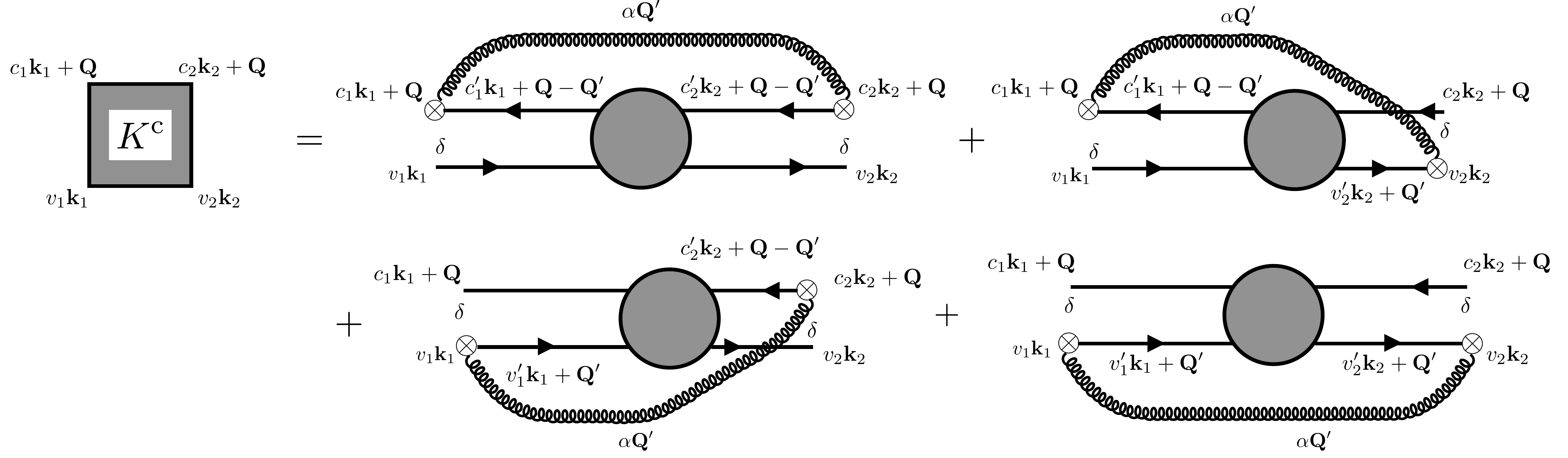}
\caption{Diagrammatic representation of the irreducible kernel to 
second order in the $e$-$ph$ interaction.}
\label{kernel}
\end{figure}

The approximations to $\S$ and $L$ discussed in 
Sections~\ref{cohxsec} and~\ref{incexsec} do not include the 
inelastic exciton-phonon scattering, responsible for exciton 
diffusion and thermalization. The inclusion of this fundamental process 
in the exciton dynamics requires an improvement of the xc function 
$L^{(v)}$. The exact $L^{(v)}$ satisfies the Bethe-Salpeter equation
\begin{align}
L^{(v)}= L^{\rm HSEX}+iL^{\rm 
HSEX}K^{\rm c}L^{(v)},
\label{wdLfinal}
\end{align}
where the the correlation kernel $K^{\rm c}$ is irreducible with 
respect to a cut of a $v$-, $D$- and $\ell$-line. 
To second-order in the {\em screened} $e$-$ph$ 
coupling we approximate 
\begin{align}
K^{\rm c,\blQ}_{c_{1}v_{1}\blk_{1},c_{2}v_{2}\blk_{2}}(z_{1},z_{2};z_{3},z_{4})&=
-\d(z_{1},z_{3})\d(z_{2},z_{4})\sum_{\a\blQ'}D_{\a\blQ'}(z_{1},z_{2})
\nn\\
\times&\Big[\sum_{c'_{1}c'_{2}}
g_{\a-\blQ'}^{s,c_{1}c'_{1}}(\blk_{1}+\blQ)g_{\a\blQ'}^{s,c'_{2}c_{2}}(\blk_{2}+\blQ-\blQ')
 L^{\blQ-\blQ'}_{c'_{1}v_{1}\blk_{1},c'_{2}v_{2}\blk_{2}}(z_{1},z_{2};z_{1},z_{2})
\nn\\-&
\sum_{c'_{1}v'_{2}}
g_{\a-\blQ'}^{s,c_{1}c'_{1}}(\blk_{1}+\blQ)g_{\a\blQ'}^{s,v_{2}v'_{2}}(\blk_{2})
L^{\blQ-\blQ'}_{c'_{1}v_{1}\blk_{1},c_{2}v'_{2}\blk_{2}+\blQ'}(z_{1},z_{2};z_{1},z_{2})
\nn\\-&
\sum_{v'_{1}c'_{2}}
g_{\a-\blQ'}^{s,v'_{1}v_{1}}(\blk_{1}+\blQ')g_{\a\blQ'}^{s,c'_{2}c_{2}}(\blk_{2}+\blQ-\blQ')
L^{\blQ-\blQ'}_{c_{1}v'_{1}\blk_{1}+\blQ',c'_{2}v_{2}\blk_{2}}
(z_{1},z_{2};z_{1},z_{2})
\nn\\+&
\sum_{v'_{1}v'_{2}}
g_{\a-\blQ'}^{s,v'_{1}v_{1}}(\blk_{1}+\blQ')g_{\a\blQ'}^{s,v_{2}v'_{2}}(\blk_{2})
L^{\blQ-\blQ'}_{c_{1}v'_{1}\blk_{1}+\blQ',c'_{2}v_{2}\blk_{2}+\blQ'}
(z_{1},z_{2};z_{1},z_{2})\Big],
\end{align}
which is represented diagrammatically in Fig.~\ref{kernel}.
The assumption of small excitation density is equivalent to 
approximate the internal Green's functions as
\begin{subequations}
\begin{align}
G_{cc\blk}(z,z')&\simeq -i\th(z,z')e^{-i\e_{c\blk}(t-t')},
\\
G_{vv\blk}(z,z')&\simeq i\th(z',z)e^{-i\e_{v\blk}(t-t')}.
\end{align}
\label{sedgf}
\end{subequations}
Consider the first 
diagram of the kernel. The structure  
$GG K^{\rm c,\blQ}$  involves 
the calculation of 
\begin{align}
\int 
dz_{1}dz_{3}G_{c_{1}\blk_{1}+\blQ}(z,z_{1})G_{c'_{1}\blk_{1}+\blQ-\blQ'}(z_{1},z')
\d(z_{3},z_{1})G_{v_{1}\blk_{1}}(z',z_{3})G_{v_{1}\blk_{1}}(z_{3},z)
\nn\\
=i G_{v_{1}\blk_{1}}(z',z)\int 
dz_{1}G_{c_{1}\blk_{1}+\blQ}(z,z_{1})G_{c'_{1}\blk_{1}+\blQ-\blQ'}(z_{1},z'),
\label{connection}
\end{align}
where we use Eqs.~(\ref{sedgf}). The gluing of two Green's 
functions allows us to recover the Feynman rules for the two-particle Green's 
function~\cite{svl-book}. It is straightforward to verify that the 
gluing argument applies to all four diagrams of the kernel. 
We emphasize that no overscreening issue arises if we use 
the screened $e$-$ph$ coupling in $K^{\rm c}$. In fact, the 
Green's functions entering an $e$-$ph$ vertex come from different xc 
functions, ensuring that screening is counted only once.
The diagrams of Fig.~\ref{kernel} have been discussed in 
Refs.~\cite{cudazzo_first-principles_2020,antonius_theory_2022,cudazzo_dynamical_2023} for the
stationary case and for $L=L^{\rm HSEX}$. 

Since $K^{\rm c}$ is already of second order in $g^{s}$ we 
approximate [see Eq.~(\ref{Lwo})] 
\begin{align}
K^{\rm c}L^{(v)}=K^{\rm 
c}\big(L+i\widetilde{L}g^{s}Dg^{s}\widetilde{L}\big)\simeq K^{\rm c} 
L,
\end{align}
which implies
\begin{align}
L\simeq L^{\rm HSEX}+iL^{\rm HSEX}K^{\rm c}[L]L-i\widetilde{L}
g^{s}Dg^{s}\widetilde{L}.
\label{Lfinal}
\end{align}
In the following we derive the equations of motion for 
all kind of excitons and show how they are coupled.

\subsection{$D$-reducible excitons}

The approximation in Eq.~(\ref{wdLfinal}) implies that, see 
Eq.~(\ref{DysonLcl}),
\begin{align}
\widetilde{L}=\widetilde{L}^{\rm SEX}+
\widetilde{L}^{\rm SEX}K^{\rm c}[L]\widetilde{L}.
\label{tildeLfinal}
\end{align}
The last term in Eq.~(\ref{Lfinal}) with $\widetilde{L}$ from 
Eq.~(\ref{tildeLfinal}) provides 
an improved approximation to the $D$-reducible exciton propagator 
$N^{(D)}$, compare with 
Eq.~(\ref{cxix}). 
The equation of motion for $N^{(D),<}$ in the irreducible 
exciton basis is identical to Eq.~(\ref{dncxixdt}) provided that we 
replace $\widetilde{N}^{\rm SEX}\to \widetilde{N}$. 
To calculate the collision integral 
we extend the MGKBA to 
$\widetilde{N}^{\lessgtr}$ along the lines outlined 
Ref.~\cite{pavlyukh_photoinduced_2021}  [compare with Eq.~(\ref{tildeN})]
\begin{align}
\widetilde{N}^{\lessgtr}_{\widetilde{\l}\blQ}(t,t')=\Big[\th(t-t')\;
\widetilde{N}^{\lessgtr}_{\widetilde{\l}\blQ}(t,t)
+\th(t'-t)\widetilde{N}^{\lessgtr}_{\widetilde{\l}\blQ}(t',t')\Big]
e^{-i\widetilde{E}_{\widetilde{\l}\blQ}(t-t')},
\label{mgkbaninc}
\end{align}
where $\widetilde{N}^{<}_{\widetilde{\l}\blQ}(t,t)=
\widetilde{N}_{\widetilde{\l}\blQ}(t)$ is 
the number of irreducible excitons, and
$\widetilde{N}^{>}_{\widetilde{\l}\blQ}(t,t)=
1+\widetilde{N}_{\widetilde{\l}\blQ}(t)$. Then, the
improved version of the equation of motion Eq.~(\ref{dnephdt}) reads 
\begin{align}
\frac{d}{dt}
N^{(D)}_{\widetilde{\l}\blQ}(t)=
\sum_{\l'}\Big[\widetilde{\G}^{\rm pol}_{\l'\widetilde{\l}\blQ}
+\widetilde{\G}_{\l'\widetilde{\l}\blQ}\widetilde{N}_{\widetilde{\l}\blQ}(t)
\Big]|\r_{\l'}(t)|^{2},
\label{dnephdt2}
\end{align}
where the rates
\begin{align}
\widetilde{\G}_{\l\widetilde{\l}'\blQ}&=
2\p\sum_{\a}\frac{|\widetilde{\callG}^{\widetilde{\l}'\l}_{\a}(\blQ)|^{2}}{2\w_{\a\blQ}}
\Big[\d(\widetilde{E}_{\widetilde{\l}'\blQ}-E_{\l\bz}+\w_{\a\blQ})-
\d(\widetilde{E}_{\widetilde{\l}'\blQ}-E_{\l\bz}-\w_{\a\blQ})\Big]
\label{gammatilde}
\end{align}
depend on the irreducible exciton-phonon coupling defined in 
Eq.~(\ref{callGx2irr}). Equation (\ref{dnephdt2}) couples the 
$D$-reducible excitons to the irreducible excitons $\widetilde{N}$ 
and the coherent excitons $|\r|^{2}$.

\subsection{Irreducible excitons}

The equation of motion for $\widetilde{N}_{\widetilde{\l}\blQ}$ can 
be derived from Eq.~(\ref{tildeLfinal}).
As the kernel is proportional to $\d(z_{1},z_{3})\d(z_{2},z_{4})$ we 
infer that this equation can be closed on 
$\widetilde{N}(z,z')=\widetilde{L}(z,z';z,z')$ and 
$N^{\rm inc}(z,z')=L(z,z';z,z')$ [see Eqs.~(\ref{dN}) and (\ref{Ntilde})], 
thus becoming an integral equation  
for two-time functions on the 
Keldysh contour. Considering only the diagonal elements of 
$\widetilde{N}$ in the irreducible exciton basis and the diagonal 
elements of $N^{\rm inc}$ in the exciton basis
 we find
\begin{subequations}
\begin{align}
\Big[i\frac{d}{dz}-\widetilde{E}_{\widetilde{\l}\blQ}\Big]	
\widetilde{N}_{\widetilde{\l}\blQ}(z,z')=i\d(z,z')+\int d\bar{z}\,
\widetilde{\P}_{\widetilde{\l}\blQ}(z,\bar{z})
\widetilde{N}_{\widetilde{\l}\blQ}(\bar{z},z'),
\label{eomntilde1}
\\
\Big[-i\frac{d}{dz'}-\widetilde{E}_{\widetilde{\l}\blQ}\Big]	
\widetilde{N}_{\widetilde{\l}\blQ}(z,z')=i\d(z,z')+\int d\bar{z}\,
\widetilde{N}_{\widetilde{\l}\blQ}(z,\bar{z})
\widetilde{\P}_{\widetilde{\l}\blQ}(\bar{z},z'),
\label{eomntilde2}
\end{align}	
\label{eomntilde}
\end{subequations}
where we use Eq.~(\ref{tildeN}) and introduce the irreducible exciton self-energy 
\begin{align}
\widetilde{\P}_{\widetilde{\l}\blQ}(z,\bar{z})\equiv 
\sum_{\l'}\sum_{\blQ'\a}
|\widetilde{\callG}^{\widetilde{\l}\l'}_{\a-\blQ'}(\blQ)|^{2}\,
N^{\rm inc}_{\l'\blQ-\blQ'}(z,\bar{z})\,D_{\a\blQ'}(z,\bar{z}).
\label{xsigma}
\end{align}
The irreducible exciton-phonon coupling in $\widetilde{\P}$ 
generalizes the one in Eq.~(\ref{callGx2irr}) to finite momentum 
transfer:
\begin{align}
\widetilde{\callG}^{\widetilde{\l}\l'}_{\a-\blQ'}(\blQ)\equiv 
\sum_{c_{1}c_{2}v_{1}\blk_{1}}
\widetilde{A}^{\widetilde{\l}\blQ\ast}_{c_{1}v_{1}\blk_{1}}g^{s,c_{1}c_{2}}_{\a-\blQ'}(\blk_{1}+\blQ)
A^{\l'\blQ-\blQ'}_{c_{2}v_{1}\blk_{1}}
-
\sum_{c_{1}v_{1}v_{2}\blk_{1}}\widetilde{A}^{\widetilde{\l}\blQ\ast}_{c_{1}v_{1}\blk_{1}}
g^{s,v_{2}v_{1}}_{\a-\blQ'}(\blk_{1}+\blQ')A^{\l'\blQ-\blQ'}_{c_{1}v_{2}\blk_{1}+\blQ'}.
\label{callGx2irrQ}
\end{align}
The Matsubara component of the irreducible exciton self-energy agrees 
with the result in Ref.~\cite{antonius_theory_2022} if we replace 
$N^{\rm inc}$ with $N^{\rm HSEX}$ and $\widetilde{\callG}$ with 
$\callG$. As we see below, having a self-energy which is 
a functional of $N^{\rm inc}$ is essential to derive  a
Boltzmann-like equation.

To extract the equation of motion for the number of irreducible excitons
we subtract
Eq.~(\ref{eomntilde2}) from Eq.~(\ref{eomntilde1}) and then 
set $z=t_{-}$ and $z'=t_{+}$. Extending the MGKBA 
in Eq.~(\ref{mgkbaninc}) to the propagator $N^{\rm inc}$, i.e.,
\begin{align}
N^{\rm inc,\lessgtr}_{\l\blQ}(t,t')=\Big[\th(t-t')\;
N^{\rm inc,\lessgtr}_{\l\blQ}(t,t)
+\th(t'-t)N^{\rm inc,\lessgtr}_{\l\blQ}(t',t')\Big]
e^{-iE_{\l\blQ}(t-t')},
\label{mgkbaninc(v)}
\end{align}
and implementing the Markov approximation we obtain
\begin{align}
\frac{d}{dt}\widetilde{N}_{\widetilde{\l}\blQ}(t)=
-\widetilde{\G}^{\rm out}_{\widetilde{\l}\blQ}(t)
\widetilde{N}_{\widetilde{\l}\blQ}(t)
+\widetilde{\G}^{\rm in}_{\widetilde{\l}\blQ}(t)
\big(1+\widetilde{N}_{\widetilde{\l}\blQ}(t)\big),
\label{dnxphdt}
\end{align}
where the irreducible excitonic rates are given by
\begin{subequations}
\begin{align}
\widetilde{\G}^{\rm out}_{\widetilde{\l}\blQ}&=2\p
\sum_{\l'\a\blQ'}
\frac{\big|\widetilde{\callG}^{\widetilde{\l}\l'}_{\a-\blQ'}(\blQ)\big|^{2}}{2\w_{\a\blQ'}}
\big(1+N^{\rm inc}_{\l'\blQ-\blQ'}\big)
\nn\\
&\times\Big[
\d\big(E_{\l'\blQ-\blQ'}-\widetilde{E}_{\widetilde{\l}\blQ}+\w_{\a\blQ'}\big)(1+f^{\rm 
ph}_{\a\blQ'})
+\d\big(E_{\l'\blQ-\blQ'}-\widetilde{E}_{\widetilde{\l}\blQ}-\w_{\a\blQ'}\big)f^{\rm ph}_{\a-\blQ'}\Big],
\\
\widetilde{\G}^{\rm in}_{\widetilde{\l}\blQ}&=2\p
\sum_{\l'}\sum_{\blQ'\a}
\frac{\big|\widetilde{\callG}^{\widetilde{\l}\l'}_{\a-\blQ'}(\blQ)\big|^{2}}{2\w_{\a\blQ'}}
N^{\rm inc}_{\l'\blQ-\blQ'}
\nn\\
&\times \Big[
\d\big(E_{\l'\blQ-\blQ'}-\widetilde{E}_{\widetilde{\l}\blQ}+\w_{\a\blQ'}\big)f^{\rm ph}_{\a\blQ'}
+\d\big(E_{\l'\blQ-\blQ'}-\widetilde{E}_{\widetilde{\l}\blQ}-\w_{\a\blQ'}\big)(1+f^{\rm 
ph}_{\a-\blQ'})\Big].
\end{align}
\label{xrates}
\end{subequations}
Notice that this is {\em not} a Boltzmann equation for 
$\widetilde{N}$ since the rates depend on the occupations $N^{\rm 
inc}$.

\subsection{Incoherent excitons}

The number of incoherent excitons is given by the sum of 
$v$-reducible and $D$-reducible excitons, see Eq.~(\ref{dN}) and 
(\ref{Lwo}):
\begin{align}
N^{\rm inc}=N^{(v)}+N^{(D)}.
\end{align}
According to the approximation in Eq.~(\ref{Lfinal}): 
\begin{align}
L^{(v)}=L^{\rm HSEX}+iL^{\rm HSEX}K^{\rm c}[L]L.
\label{deltaL2}
\end{align}

The equation of motion for $N^{(v)}(t)$
can be derived from Eq.~(\ref{deltaL2}),
by following the same logic as for $\widetilde{N}$. 
Considering only the diagonal elements of 
$N^{\rm inc}$  
in the excitonic basis 
we find
\begin{align}
\frac{d}{dt} N^{(v)}_{\l\blQ}(t)=
-\G^{\rm out}_{\l\blQ}(t)
N^{\rm inc}_{\l\blQ}(t)
+\G^{\rm in}_{\l\blQ}(t)
\big(1+N^{\rm inc}_{\l\blQ}(t)\big),
\label{dDnxphdt}
\end{align}
where the  excitonic rates 
\begin{subequations}
\begin{align}
\G^{\rm out}_{\l\blQ}&=2\p
\sum_{\l'\a\blQ'}
\frac{\big|\callG^{\l\l'}_{\a-\blQ'}(\blQ)\big|^{2}}{2\w_{\a\blQ'}}
\big(1+N^{\rm inc}_{\l'\blQ-\blQ'}\big)
\nn\\
&\times\Big[
\d\big(E_{\l'\blQ-\blQ'}-E_{\l\blQ}+\w_{\a\blQ'}\big)(1+f^{\rm 
ph}_{\a\blQ'})
+\d\big(E_{\l'\blQ-\blQ'}-E_{\l\blQ}-\w_{\a\blQ'}\big)f^{\rm ph}_{\a-\blQ'}\Big],
\\
\G^{\rm in}_{\l\blQ}&=2\p
\sum_{\l'}\sum_{\blQ'\a}
\frac{\big|\callG^{\l\l'}_{\a-\blQ'}(\blQ)\big|^{2}}{2\w_{\a\blQ'}}
N^{\rm inc}_{\l'\blQ-\blQ'}
\nn\\
&\times \Big[
\d\big(E_{\l'\blQ-\blQ'}-E_{\l\blQ}+\w_{\a\blQ'}\big)f^{\rm ph}_{\a\blQ'}
+\d\big(E_{\l'\blQ-\blQ'}-E_{\l\blQ}-\w_{\a\blQ'}\big)(1+f^{\rm 
ph}_{\a-\blQ'})\Big],
\end{align}
\label{xratesn0}
\end{subequations}
have the same mathematical form as the irreducible excitonic rates in 
Eq.~(\ref{xrates}), the difference being that the irreducible 
exciton energies and wavefunctions are replaced by the exciton ones. 
Notice the emergence of the original exciton-phonon coupling defined in 
Eq.~(\ref{callGx2}).

\subsection{Coherent excitons}

The improved approximation to $\widetilde{L}$ leads to an improved 
equation of motion for the exciton polarization as well. In fact, the 
collision integral $S_{\l}$ must now be evaluated with  $\widetilde{N}$, 
whose lesser component is, in general, nonzero. 
Using the MGKBA in 
Eq.~(\ref{mgkbaninc}) it is straightforward to derive 
\begin{align}
\frac{d}{dt}\r_{\l}(t)=-iE_{\l \bz}\r_{\l}(t)
-i\W_{\l}(t)- \frac{1}{2}\sum_{\widetilde{\l}'\blQ}
\Big[\widetilde{\G}^{{\rm pol}}_{\l\widetilde{\l}'\blQ}
+\widetilde{\G}_{\l\widetilde{\l}'\blQ}
\widetilde{N}_{\widetilde{\l}'\blQ}(t)
\Big]
\r_{\l}(t),
\label{drldtimp}
\end{align}
which should be compared with Eq.~(\ref{drldt}).

\section{Excitonic Bloch  equations}
\label{SBEEsec}

\begin{table}[t]
\begin{center}	
{\footnotesize    
\begin{tabular}{|c|c|}
        \hline
	 & \\
Definition  & Equations   \\
 & \\
\hline
& \\
Exciton polarization & 
$\frac{d}{dt}\r_{\l}=-iE_{\l \bz}\r_{\l}
-i\W_{\l}- \frac{1}{2}\sum_{\widetilde{\l}'\blQ}
\Big[\widetilde{\G}^{{\rm pol}}_{\l\widetilde{\l}'\blQ}
+\widetilde{\G}_{\l\widetilde{\l}'\blQ}
\widetilde{N}_{\widetilde{\l}'\blQ}
\Big]
\r_{\l}$ \\
& \\
\hline
& \\
Irreducible excitons &
$\frac{d}{dt}\widetilde{N}_{\widetilde{\l}\blQ}=
-\widetilde{\G}^{\rm out}_{\widetilde{\l}\blQ}[N^{\rm inc}]\,
\widetilde{N}_{\widetilde{\l}\blQ}
+\widetilde{\G}^{\rm in}_{\widetilde{\l}\blQ}[N^{\rm inc}]\,
\big(1+\widetilde{N}_{\widetilde{\l}\blQ}\big)$
\\
& \\
\hline
& \\
Incoherent excitons & $
\frac{d}{dt}N^{\rm inc}_{\l\blQ}=
-\G^{\rm out}_{\l\blQ}[N^{\rm inc}]\,
N^{\rm inc}_{\l\blQ}
+\G^{\rm in}_{\l\blQ}[N^{\rm inc}]\,
\big(1+N^{\rm inc}_{\l\blQ}\big)
+\sum_{\l'\widetilde{\l}}|S^{\blQ}_{\l\widetilde{\l}}|^{2}
\Big[\widetilde{\G}^{\rm pol}_{\l'\widetilde{\l}\blQ}
+\widetilde{\G}_{\l'\widetilde{\l}\blQ}\widetilde{N}_{\widetilde{\l}\blQ}
\Big]|\r_{\l'}|^{2}
$ \\
& \\
\hline
\end{tabular}
\caption{Excitonic Bloch  equations. }
\label{sbxe}
}
\end{center}
\end{table}

In Table~\ref{sbxe}, we summarize the main results from 
Section~\ref{xxsec}, highlighting the types of excitons on which the 
various rates depend. We refer to the equations in Table~\ref{sbxe} 
as the {\em excitonic Bloch  equations} (XBE).
It is noteworthy that the irreducible 
exciton-phonon coupling $\widetilde{\callG}$ governs the dynamics of 
coherent excitons, while the reducible 
exciton-phonon coupling $\callG$ governs the dynamics of incoherent 
excitons. The XBE reconciles previous works that advocate for either 
$\widetilde{\callG}$~\cite{paleari_exciton-phonon_2022} or 
$\callG$~\cite{thranhardt_quantum_2000,jiang_exciton-phonon_2007,chen_exciton-phonon_2020,cudazzo_first-principles_2020,antonius_theory_2022}, 
by clarifying that both are essential, though relevant in different 
regimes (or equivalently at different timescales).

Choosing the time origin earlier than the switch-on time of the 
external driving fields (hence $\W_{\l}(t<0)=0$), the XBE must be 
solved with initial conditions 
$\r_{\l}(0)=\widetilde{N}_{\widetilde{\l}\blQ}(0)$\linebreak$=N^{\rm 
inc}_{\l\blQ}(0)=0$.
The first line contains the equation of motion  of the exciton 
polarization, see Eq.~(\ref{drldtimp}). The quantity
\begin{align}
\g_{\l}\equiv \sum_{\widetilde{\l}'\blQ}
\Big[\widetilde{\G}^{{\rm pol}}_{\l\widetilde{\l}'\blQ}
+\widetilde{\G}_{\l\widetilde{\l}'\blQ}
\widetilde{N}_{\widetilde{\l}'\blQ}
\Big]
\end{align}
gives the exciton linewidth of a photoabsorption spectrum, and agrees 
with Ref.~\cite{paleari_exciton-phonon_2022} for infinitesimally small 
excitation densities (i.e., $\widetilde{N}=0$).
An increasing number of 
irreducible excitons accelerates the transition toward the incoherent 
regime, in agreement with the fact that the polarization lifetime  
decreases with the excitation 
density~\cite{calati_ultrafast_2021,perfetto_real-time_2023}. 
The equation of motion for the number of irreducible 
excitons, see Eq.~(\ref{dnxphdt}), is shown in the second line. 
The third line contains the equation of motion for the number of 
incoherent excitons, obtained by adding Eqs.~(\ref{dnephdt2}) and 
(\ref{dDnxphdt}). Coherent excitons are first converted 
into $D$-reducible excitons $N^{(D)}$, see Eq.~(\ref{dnephdt2}), 
which then diffuse and becomes incoherent excitons. 
Aside from the overscreening issue, the equation of motion for 
$N^{\rm inc}$ agrees with findings from the cluster 
expansion~\cite{thranhardt_quantum_2000,selig_exciton_2016,brem_exciton_2018}  
in the limit of infinitesimally small excitation densities.
This limit corresponds to setting $\widetilde{N}=0$ and retaining only 
terms linear in $N^{\rm inc}$.
In the incoherent regime, i.e., $\r_{\l}=0$, the equation of motion 
for $N^{\rm inc}$ also agrees with findings from 
Ref.~\cite{chen_first-principles_2022}. We observe, however, that 
setting $\r_{\l}=0$ from the outset yields a homogeneous equation, 
meaning that the initial value of the incoherent 
exciton population 
must be determined by other means.
The many-body diagrammatic treatment of the first principles Hamiltonian not only 
provides a many-body justification of the work in 
Ref.~\cite{chen_first-principles_2022}, but also extends it to the 
coherent regime. This extension allows for monitoring the exciton 
dynamics from the moment the optical field drives the system out of 
equilibrium.   

As $\lim_{t\to\iif}\r_{\l}(t)= 0$, 
the number of incoherent excitons  
approaches a Bose-Einstein distribution with the same temperature as 
the phonon bath -- it is straightforward to show that
the right hand side of the third XBE in Table~\ref{sbxe} 
vanishes in this scenario. This implies that $\lim_{t\to\iif} \widetilde{N}(t)$ 
follows a Bose-Einstein distribution as well. Indeed, the right hand side 
of the second XBE in Table~\ref{sbxe} vanishes if $f^{\rm ph}$, 
$N^{\rm inc}$ and $\widetilde{N}$ are all described by a 
Bose-Einstein distribution at the same temperature.

To summarize, the XBE have the merits of being 
overscreening free and applicable to nonequilibrium systems beyond 
the linear response regime. Furthermore, they preserve 
Eq.~(\ref{dNtotdt}), according to which the total number of excitons 
remains constant when the driving field is off.

\section{Conclusions}
\label{consec}

Starting from the ab initio Hamiltonian for electrons and 
phonons~\cite{stefanucci_in-and-out_2023} and using 
the many-body diagrammatic Green's function theory~\cite{svl-book} we have derived 
a first principles scheme for material 
specific predictions of nonequilibrium excitons. 
The XBE are a system of nonlinear differential equations for 
the coupled dynamics of coherent excitons, irreducible excitons and 
incoherent excitons. They encompass the initial transient regime, driven by 
external optical fields, as well as the evolution from the coherent 
to the incoherent regime, governed by exciton-phonon scatterings. 
Importantly, the XBE are
{\em free} of  overscreening issues. 
It is worth remarking that the first-principles formulation
presented in this work involves only screened $e$-$ph$ couplings. In 
Appendix~\ref{gbareapp} we outline an alternative first-principles 
formulation that eliminates the need for introducing irreducible 
excitons but involves the bare $e$-$ph$ coupling.
In the ab initio theory of electrons and phonons the bare $g$ 
appears in the exact formula of the phonon 
self-energy~\cite{marini_many-body_2015,feliciano_electron-phonon_2017,marini_equilibrium_2023} 
as well as in the coupling to coherent 
phonons~\cite{stefanucci_in-and-out_2023,stefanucci_semiconductor_2024,perfetto_theory_2024}.
However, in the present context  the use of a bare $g$ 
complicates the formulation since both intraband and interband 
$e$-$ph$ coupling must be accounted for. In fact, the dressing 
of $g^{cc'}$ and $g^{vv'}$ is mainly due to $g^{cv}$ and $g^{vc}$.

The XBE form a minimal set of 
equations for describing exciton formation, diffusion and 
thermalization for not too high excitation densities. They can be 
improved along several directions that we wish to discuss here. 
First, the lifting of the 
Markov approximation through the introduction of higher order correlators. This 
idea has been already implemented in nonequilibrium systems of
electrons~\cite{schlunzen_achieving_2020,joost_g1-g2_2020,pavlyukh_photoinduced_2021,pavlyukh_time-linear_2022,perfetto_real_2022} and 
bosons~\cite{karlsson_fast_2021,pavlyukh_interacting_2022,perfetto_real-time_2023,pavlyukh_cheers_2023} 
for several many-body approximations, and it is also possible to implement it in this context.  
Second, the extension to nonequilibrium phonons. In this 
work, we have assumed that phonons remain in thermal equilibrium. 
However, at sufficiently high excitation densities, this assumption 
becomes unrealistic. The equation of motion for the phonon 
occupations can be derived following the approach outlined in 
Ref.~\cite{stefanucci_semiconductor_2024}. However, the phononic 
self-energy must be refined to account for excitonic effects in the 
polarization. Third, at large enough excitation density the Coulomb 
mediated exciton-exciton scattering can no longer be ignored. This 
interaction has been investigated in 
Refs.~\cite{sun_stimulated_2000,shahnazaryan_exciton-exciton_2017,erkensten_exciton-exciton_2021},  
and it gives rise to 
additional rates in the equations of motion.
Fourth, excitons can strongly couple to coherent optical phonons. 
This interaction is governed by the bare $e$-$ph$ 
coupling~\cite{stefanucci_in-and-out_2023,perfetto_theory_2024} and 
is responsible for a time-dependent shift of the exciton energies. 
Given the long timescale of optical nuclear displacements, 
we expect that an adiabatic approximation of the XBE, 
i.e., $E_{\lambda\mathbf{Q}} \to 
E_{\lambda\mathbf{Q}}(t)$, is reasonably accurate. Fifth, the 
inclusion of exciton recombination through the quantized treatment of 
photons. This aspect has been already covered in, e.g., 
Ref.~\cite{thranhardt_quantum_2000}, and does not present 
criticalities as the ab initio Hamiltonian for electrons and photons 
has long been  well established.

We hope that our contribution can clarify the pitfalls
inherent in model Hamiltonians, serve as a solid ground for the 
development of a rigorous theory 
of  nonequilibrium excitons, phonons and photons, and inspire 
parameter-free numerical schemes for real-time simulations of 
excitonic materials.

\section*{Acknowledgements}


\paragraph{Funding information}
This work has been supported by MIUR PRIN (Grant No. 2022WZ8LME),
INFN through the TIME2QUEST project, 
Tor Vergata University through Project TESLA, and the European Union’s Horizon
Europe research and innovation programme under the Marie
Sk{\polishl}odowska-Curie Doctoral Networks grant agreement No. 101118915 –
TIMES.

\begin{appendix}

\section{On the polarization rates}
\label{polrates}

For the evaluation of Eq.~(\ref{xphscatt}) we assume that the 
Green's function $G_{\m\m'\blk}$ with band indices $\m,\m'$ either 
both conduction or both valence is diagonal. Let us analyze 
$\big[\S^{\rm FM}(z,\bar{z})G(\bar{z},z^{+})\big]_{cv\blk}$.
Focusing solely on the dependence on the electronic band indices, this 
term contains either 
$g^{s,cc_{1}}G_{c_{1}c_{1}}(z,\bar{z})g^{s,c_{1}c_{2}}G_{c_{2}v}(\bar{z},z^{+})$ or 
$g^{s,cc_{1}}G_{c_{1}v_{1}}(z,\bar{z})g^{s,v_{1}v}G_{vv}(\bar{z},z^{+})$.
The second diagram in Fig.~\ref{Sigma-el} contributes as 
$\big[\S^{(1)}(z,\bar{z})G(\bar{z},z^{+})\big]_{cv\blk}$. 
We have the following possible structures 
\begin{subequations}
\begin{align}
g^{s,cc_{1}}G_{c_{1}c_{1}}(z,\bar{z})G_{c_{2}c_{2}}(\bar{z}',\bar{z}'')
g^{s,c_{2}c_{3}}G_{c_{3}v_{1}}(\bar{z}'',\bar{z}')G_{vv}(\bar{z},z),
\\
g^{s,cc_{1}}G_{c_{1}v_{1}}(z,\bar{z})G_{v_{2}c_{2}}(\bar{z}',\bar{z}'')
g^{s,c_{2}c_{3}}G_{c_{3}c_{3}}(\bar{z}'',\bar{z}')G_{c_{4}v}(\bar{z},z),
\\
g^{s,cc_{1}}G_{c_{1}c_{1}}(z,\bar{z})G_{c_{2}v_{1}}(\bar{z}',\bar{z}'')
g^{s,v_{1}v_{2}}G_{v_{2}v_{2}}(\bar{z}'',\bar{z}')G_{vv}(\bar{z},z),
\\
g^{s,cc_{1}}G_{c_{1}v_{1}}(z,\bar{z})G_{v_{2}v_{2}}(\bar{z}',\bar{z}'')
g^{s,v_{2}v_{3}}G_{v_{3}c_{2}}(\bar{z}'',\bar{z}')G_{c_{3}v}(\bar{z},z).
\end{align}
\end{subequations}
We see that the first and third combinations are linear in the off-diagonal 
$G$ whereas the second and third combinations are cubic, and can 
therefore be ignored. We finally consider 
$\big[\S^{(2)}(z,\bar{z})G(\bar{z},z^{+})\big]_{cv\blk}$. We 
have the following possible structure
\begin{subequations}
\begin{align}
G_{c_{1}c_{1}}(\bar{z}',\bar{z}'')g^{c_{1}c_{2}}G_{c_{2}v_{2}}(\bar{z}'',\bar{z}')	
G_{v_{3}c_{3}}(z,\bar{z})g^{c_{3}c_{4}}G_{c_{4}v}(\bar{z},z),
\\
G_{c_{1}c_{1}}(\bar{z}',\bar{z}'')g^{c_{1}c_{2}}G_{c_{2}v_{2}}(\bar{z}'',\bar{z}')	
G_{v_{3}v_{3}}(z,\bar{z})g^{v_{3}v}G_{vv}(\bar{z},z),
\\
G_{c_{1}v_{1}}(\bar{z}',\bar{z}'')g^{v_{1}v_{2}}G_{v_{2}v_{2}}(\bar{z}'',\bar{z}')	
G_{v_{3}c_{3}}(z,\bar{z})g^{c_{3}c_{4}}G_{c_{4}v}(\bar{z},z),
\\
G_{c_{1}v_{1}}(\bar{z}',\bar{z}'')g^{v_{1}v_{2}}G_{v_{2}v_{2}}(\bar{z}'',\bar{z}')	
G_{v_{3}v_{3}}(z,\bar{z})g^{v_{3}v}G_{vv}(\bar{z},z).
\end{align}
\end{subequations}
The first and third combinations are cubic in the off-diagonal $G$, and we 
discard them. The second and fourth combinations contain 
$G_{v_{3}v_{3}}(z,\bar{z})G_{vv}(\bar{z},z)$ which scales linearly 
with the excitation density, and we therefore discard these 
combinations as well.
We can use the ansatz in Eqs.~(\ref{vcansatz}) 
to express the off-diagonal elements of $\widetilde{L}$ in terms of the 
irreducible exciton propagator:
\begin{subequations}
\begin{align}
&\widetilde{L}^{\blQ}_{cv\blk,c_{1}c_{2}\blk_{1}}(z,\bar{z};z,\bar{z})	=
\sum_{v_{1}}\widetilde{N}^{\blQ}_{cv\blk,c_{1}v_{1}\blk_{1}}(z,\bar{z})
\r_{c_{2}v_{1}\blk_{1}}(\bar{t}),
\\
&\widetilde{L}^{\blQ}_{cv\blk,v_{2}v_{1}\blk_{1}}(z,\bar{z};z,\bar{z})=
-\sum_{c_{1}}\widetilde{N}^{\blQ}_{cv\blk,c_{1}v_{1}\blk_{1}}(z,\bar{z})
\r_{c_{1}v_{2}\blk_{1}+\blQ}(\bar{t}).
\end{align}
\label{odNee}
\end{subequations}
After some algebra we find
\begin{align}
\big[\S(z,\bar{z})G(\bar{z},z^{+})\big]_{cv\blk}&=
i\sum_{c_{1}c_{2}c_{3}v_{1},\a\blQ}
g^{s,cc_{1}}_{\a-\blQ}(\blk)\widetilde{N}^{-\blQ}_{c_{1}v\blk,c_{2}v_{1}\blk_{1}}(z,\bar{z})
\r_{c_{3}v_{1}\blk_{1}}(\bar{t})g^{s,c_{2}c_{3}}_{\a\blQ}(\blk_{1}-\blQ)D_{\a\blQ}(z,\bar{z})
\nn\\
&-i\sum_{c_{1}c_{2}v_{1}v_{2},\a\blQ}
g^{s,cc_{1}}_{\a-\blQ}(\blk)
\widetilde{N}^{-\blQ}_{c_{1}v\blk,c_{2}v_{2}\blk_{1}}(z,\bar{z})\r_{c_{2}v_{1}\blk_{1}-\blQ}(\bar{t})
g^{s,v_{1}v_{2}}_{\a\blQ}(\blk_{1}-\blQ)D_{\a\blQ}(z,\bar{z}).
\end{align}
We now contract the left hand side with the exciton wave functions, 
and expand $\widetilde{N}$ in the irreducible excitonic basis 
and $\r$ in the excitonic basis
\begin{align}
\sum_{cv\blk}A^{\l\bz\ast}_{cv\blk}
\big[\S(z,\bar{z})G(\bar{z},z^{+})\big]_{cv\blk}
&=i\sum_{\widetilde{\l}_{1}\l_{2}\a\blQ}
\left(\sum_{cc_{1}v\blk}A^{\l\bz\ast}_{cv\blk}g^{s,cc_{1}}_{\a-\blQ}(\blk)
\widetilde{A}^{\widetilde{\l}_{1}-\blQ}_{c_{1}v\blk}\right)
\widetilde{N}_{\widetilde{\l}_{1}-\blQ}(z,\bar{z})
\r_{\l_{2}}(\bar{t})D_{\a\blQ}(z,\bar{z})
\nn\\
&\times \left(\sum_{c_{2}c_{3}v_{2}\blk_{1}}
\widetilde{A}^{\widetilde{\l}_{1}-\blQ\ast}_{c_{2}v_{1}\blk_{1}}
g^{s,c_{2}c_{3}}_{\a\blQ}(\blk_{1}-\blQ)
A^{\l_{2}\bz}_{c_{3}v_{1}\blk_{1}}\right)
\nn\\
&-i\sum_{\widetilde{\l}_{1}\l_{2}\a\blQ}
\left(\sum_{cc_{1}v\blk}A^{\l\bz\ast}_{cv\blk}g^{s,cc_{1}}_{\a-\blQ}(\blk)
\widetilde{A}^{\widetilde{\l}_{1}-\blQ}_{c_{1}v\blk}\right)
\widetilde{N}_{\widetilde{\l}_{1}-\blQ}(z,\bar{z}) \r_{\l_{2}}(\bar{t})D_{\a\blQ}(z,\bar{z})
\nn\\
&\times\left(\sum_{c_{2}v_{2}v_{3}\blk_{1}}
\widetilde{A}^{\widetilde{\l}_{1}-\blQ\ast}_{c_{2}v_{2}\blk_{1}}
g^{s,v_{3}v_{2}}_{\a\blQ}(\blk_{1}-\blQ)
A^{\l_{2}\bz}_{c_{2}v_{3}\blk_{1}-\blQ}\right).
\label{AsigmaG}
\end{align}	
It is useful to define the following quantities
\begin{subequations}
\begin{align}
\sum_{c_{1}c_{2}v_{1}\blk_{1}}
\widetilde{A}^{\widetilde{\l}\blQ\ast}_{c_{1}v_{1}\blk_{1}}
g^{s,c_{1}c_{2}}_{\a-\blQ'}(\blk_{1}+\blQ)
A^{\l'\blQ-\blQ'}_{c_{2}v_{1}\blk_{1}}&\equiv 
\widetilde{g}^{(c)\widetilde{\l}\l'}_{\a-\blQ'}(\blQ),
\\
\sum_{c_{1}c_{2}v_{1}\blk_{1}}
A^{\l\blQ-\blQ'\ast}_{c_{1}v_{1}\blk_{1}}g^{s,c_{1}c_{2}}_{\a\blQ'}(\blk_{1}+\blQ-\blQ')
\widetilde{A}^{\widetilde{\l}'\blQ}_{c_{2}v_{1}\blk_{1}}&=
\widetilde{g}^{(c)\widetilde{\l}'\l\ast}_{\a-\blQ'}(\blQ),
\\
\sum_{c_{1}v_{1}v_{2}\blk_{1}}\widetilde{A}^{\widetilde{\l}\blQ\ast}_{c_{1}v_{1}\blk_{1}}
g^{s,v_{2}v_{1}}_{\a-\blQ'}(\blk_{1}+\blQ')A^{\l'\blQ-\blQ'}_{c_{1}v_{2}\blk_{1}+\blQ'}
&\equiv \widetilde{g}^{(v)\widetilde{\l}\l'}_{\a-\blQ'}(\blQ),
\\
\sum_{c_{1}v_{1}v_{2}\blk_{1}}A^{\l\blQ-\blQ'\ast}_{c_{1}v_{1}\blk_{1}+\blQ'}
g^{s,v_{2}v_{1}}_{\a\blQ'}(\blk_{1})\widetilde{A}^{\widetilde{\l}'\blQ}_{c_{1}v_{2}\blk_{1}}
&= \widetilde{g}^{(v)\widetilde{\l}'\l\ast}_{\a-\blQ'}(\blQ),
\end{align}
\label{xphcoupl}
\end{subequations}
where we use the property 
$g^{\m\n}_{\a-\blQ}(\blk)=g^{\n\m\ast}_{\a\blQ}(\blk-\blQ)$, see 
Eq.~(\ref{ephcoupprop}). Then we can rewrite Eq.~(\ref{AsigmaG}) as 
\begin{align}
\sum_{cv\blk}A^{\l\bz\ast}_{cv\blk}
\big[\S(z,\bar{z})G(\bar{z},z^{+})\big]_{cv\blk}&=
i\sum_{\widetilde{\l}_{1}\l_{2}}\sum_{\a\blQ}
\widetilde{g}^{(c)\widetilde{\l}_{1}\l\ast}_{\a\blQ}(-\blQ)
\widetilde{N}_{\widetilde{\l}_{1}-\blQ}(z,\bar{z})
\nn\\
&\times\Big[\widetilde{g}^{(c)\widetilde{\l}_{1}\l_{2}}_{\a\blQ}(-\blQ)-
\widetilde{g}^{(v)\widetilde{\l}_{1}\l_{2}}_{\a\blQ}(-\blQ)\Big]
\r_{\l_{2}}(\bar{t})D_{\a\blQ}(z,\bar{z}).
\label{SG}
\end{align}
Proceeding along the same lines we can show that
\begin{align}
\sum_{cv\blk}A^{\l\bz\ast}_{cv\blk}
\big[G(z,\bar{z})\S(\bar{z},z^{+})\big]_{cv\blk}&=
i\sum_{\widetilde{\l}_{1}\l_{2}}\sum_{\a\blQ}
\widetilde{g}^{(v)\widetilde{\l}_{1}\l\ast}_{\a\blQ}(-\blQ)
\widetilde{N}_{\widetilde{\l}_{1}-\blQ}(z,\bar{z})
\nn\\
&\times \Big[\widetilde{g}^{(c)\widetilde{\l}_{1}\l_{2}}_{\a\blQ}(-\blQ)-
\widetilde{g}^{(v)\widetilde{\l}_{1}\l_{2}}_{\a\blQ}(-\blQ)\Big]
\r_{\l_{2}}(\bar{t})D_{\a\blQ}(z,\bar{z}).
\label{GS}
\end{align}
The collision integral $S_{\l}$ is the difference between Eqs.~(\ref{SG}) and 
(\ref{GS}). Using the definition of the irreducible exciton-phonon 
coupling in Eq.~(\ref{callGx2irr}) we conclude that
\begin{align}
S_{\l}(t)
=i\sum_{\widetilde{\l}_{1}\l_{2}}\sum_{\a\blQ}\int\!\! d\bar{z}\;
\widetilde{\callG}^{\widetilde{\l}_{1}\l\ast}_{\a}(\blQ)
\widetilde{N}_{\widetilde{\l}_{1}\blQ}(z,\bar{z})
\widetilde{\callG}^{\widetilde{\l}_{1}\l_{2}}_{\a}(\blQ)
\r_{\l_{2}}(\bar{t})D_{\a-\blQ}(z,\bar{z}),
\label{slambda2}
\end{align}
where we rename $\blQ\to-\blQ$.

We now observe that for any Keldysh function $k(z,z')$
\begin{align}
\int d\bar{z} \,k(z,\bar{z})=\int d\bar{t}\, k^{\rm R}(t,\bar{t}).
\end{align}
Taking into account that for the product of functions~\cite{svl-book}
\begin{align}
[\widetilde{N}D]^{\rm R}(t,\bar{t})=
\widetilde{N}^{\rm R}(t,\bar{t})D^{>}(t,\bar{t})+
\widetilde{N}^{<}(t,\bar{t})D^{\rm 
R}(t,\bar{t}),
\end{align}
we can rewrite Eq.~(\ref{slambda2}) as
\begin{align}
S_{\l}(t)
&=i\sum_{\widetilde{\l}_{1}\l_{2}}
\sum_{\a\blQ}\widetilde{\callG}^{\widetilde{\l}_{1}\l\ast}_{\a}(\blQ)
\widetilde{\callG}^{\widetilde{\l}_{1}\l_{2}}_{\a}(\blQ)
\nn\\
&\times \int^{t}\!\! d\bar{t}\;
\big[\widetilde{N}^{\rm R}_{\widetilde{\l}_{1}\blQ}(t,\bar{t})
D^{<}_{\a-\blQ}(t,\bar{t})+
\widetilde{N}^{<}_{\widetilde{\l}_{1}\blQ}(t,\bar{t})
D^{\rm R}_{\a-\blQ}(t,\bar{t})\big]
\r_{\l_{2}}(\bar{t}).
\label{slambda2}
\end{align}
To evaluate this quantity we use the MGKBA in Eq.~(\ref{mgkbaninc}) 
for $\widetilde{N}$, the dressed $D$ of the Born-Oppenheimer 
approximation~\cite{stefanucci_semiconductor_2024}
\begin{align}
D^{\lessgtr}_{\a\blQ}(t,t')&=\frac{\th(t-t')}{2i\w_{\a\blQ}}
\big[n^{\lessgtr}_{\a\blQ}(t)e^{-i\w_{\a\blQ}(t-t')}+
n^{\gtrless}_{\a-\blQ}(t)e^{i\w_{\a\blQ}(t-t')}\big]
\nn\\
&+\frac{\th(t'-t)}{2i\w_{\a\blQ}}
\big[n^{\lessgtr}_{\a\blQ}(t')e^{-i\w_{\a\blQ}(t-t')}+
n^{\gtrless}_{\a-\blQ}(t')
e^{i\w_{\a\blQ}(t-t')}\big],
\label{D11<gkba}
\end{align}
where $n^{<}_{\a\blQ}=f^{\rm ph}_{\a\blQ}$ and 
$n^{>}_{\blQ\a}(t)=f^{\rm ph}_{\a\blQ}+1$,
and approximate 
\begin{align}
\r_{\l}(t')=e^{-iE_{\l\bz}(t'-t)}\r_{\l}(t),
\label{polansatz}
\end{align}
see Eq.~(\ref{drldt}).
A technical remark. The self-energy in Fig.~\ref{Sigma-el} {\em is not 
positive definite} in the sense of 
Ref.~\cite{stefanucci_diagrammatic_2014}, and it gives rise 
to complex rates for nonvanishing phononic coherences
$\Theta_{\a\blQ}=$ $\Theta_{\a-\blQ}(t)= \bra
\hat{b}_{\a\blQ}\hat{b}_{\a-\blQ}\ket$. 
We therefore discard phononic coherence in our treatment. 
Taking the Markovian limit we find
\begin{align}
S_{\l}(t)=\frac{1}{2}\sum_{\widetilde{\l}'\l''\blQ}
\Big[\widetilde{\G}^{{\rm pol}}_{\l\widetilde{\l}'\l''\blQ}(t)+
\widetilde{\G}_{\l\widetilde{\l}'\l''\blQ}(t)
\widetilde{N}_{\widetilde{\l}'\blQ}(t)\Big]
\r_{\l''}(t),
\label{slambda2}
\end{align}
where 
\begin{subequations}
\begin{align}
\widetilde{\G}^{\rm pol}_{\l\widetilde{\l}'\l''\blQ}&=
2\p\sum_{\a}\frac{
\widetilde{\callG}^{\widetilde{\l}'\l\ast}_{\a}(\blQ)
\widetilde{\callG}^{\widetilde{\l}'\l''\ast}_{\a}(\blQ)}{2\w_{\a\blQ}}
\Big[\d(\widetilde{E}_{\widetilde{\l}'\blQ}-E_{\l''\bz}+\w_{\a\blQ})n^{>}_{\a-\blQ}
+
\d(\widetilde{E}_{\widetilde{\l}'\blQ}-E_{\l''\bz}-\w_{\a\blQ})
n^{<}_{\a\blQ}\Big],
\label{gammacoh2}
\\
\widetilde{\G}_{\l\widetilde{\l}'\l''\blQ}&=
2\p\sum_{\a}\frac{
\widetilde{\callG}^{\widetilde{\l}'\l\ast}_{\a}(\blQ)
\widetilde{\callG}^{\widetilde{\l}'\l''\ast}_{\a}(\blQ)}{2\w_{\a\blQ}}
\Big[\d(\widetilde{E}_{\widetilde{\l}'\blQ}-E_{\l''\bz}+\w_{\a\blQ})-
\d(\widetilde{E}_{\widetilde{\l}'\blQ}-E_{\l''\bz}-\w_{\a\blQ})\Big].
\label{gammainc2}
\end{align}
\label{gamma2}
\end{subequations}
The result reduces to Eqs.~(\ref{gamma}) and (\ref{gammatilde})
when retaining only 
the contribution $\l''=\l$.

\section{Irreducible excitons or bare $e$-$ph$ coupling?}
\label{gbareapp}

The challenge in formulating the theory using excitons instead of 
irreducible excitons lies in the appearance of the product of bare 
and screened $e$-$ph$ couplings in $\S G= 
g^{s}DgL^{\rm HSEX}$, see discussion below Eq.~(\ref{oscalc}). Since 
dynamical effects in the dressing of $g$ are typically disregarded, 
the Markov limit would result in non-positive rates. One possible 
strategy to formulate the theory exclusively in terms of excitons is 
to work with bare $e$-$ph$ couplings. 

\begin{figure}[tbp]
    \centering
\includegraphics[width=0.7\textwidth]{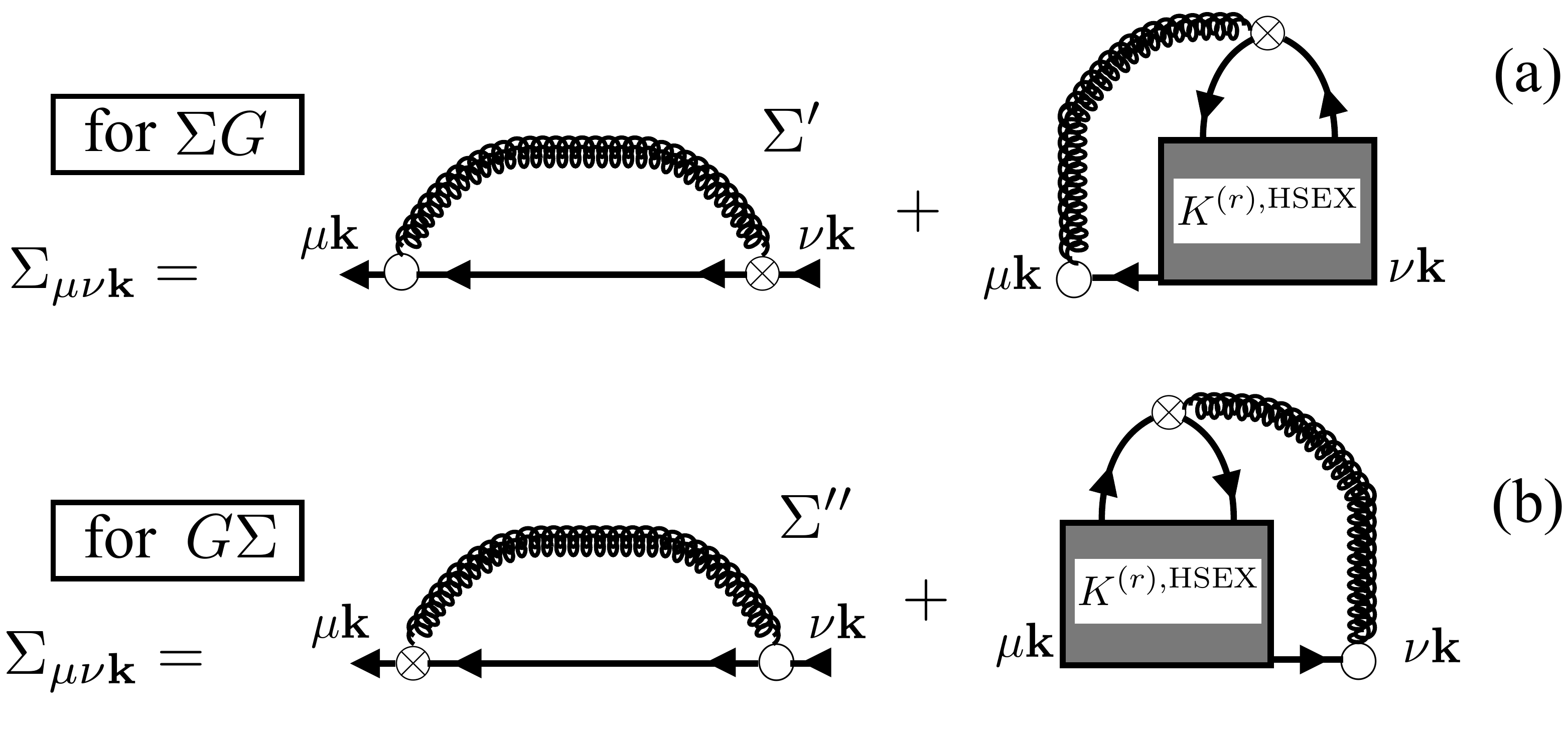}
\caption{Overscreening-free approximation of the phononic self-energy 
leading to a formulation in terms of bare $e$-$ph$ couplings. }
\label{SGGS}
\end{figure}

In Appendix~\ref{polrates} we have shown that for small excitation 
densities only $\S^{\rm FM}+\S^{(1)}$ contributes 
to $[\S G]_{cv\blk}$, and only $\S^{\rm FM}+\S^{(2)}$ contributes 
to $[G\S]_{cv\blk}$. Let us slightly modify the self-energy. For  
$[\S G]_{cv\blk}$ we use $\S'=igDGg^{s}$, instead of 
$\S^{\rm FM}=ig^{s}DGg^{s}$, and evaluate $\S^{(1)}$ with a 
bare $g$ on the left (keeping the 
screened $g^{s}$ at the top), see Fig.~\ref{SGGS}(a). Then 
\begin{align}
\S G = igD\ell g^{s}+igD(\widetilde{L}^{\rm SEX}-\ell)g^{s}=
igD\widetilde{L}^{\rm SEX}(1-ivL^{\rm HSEX})g=igDL^{\rm HSEX}g,
\end{align}
where in the last equality we use Eq.~(\ref{DysonLcl}).
Similarly, for $[G\S]_{cv\blk}$ we use $\S''=ig^{s}GDg$ , instead of 
$\S^{\rm FM}=ig^{s}DGg^{s}$,
and evaluate $\S^{(2)}$ with a bare $g$ on the 
right (keeping the 
screened $g^{s}$ at the top). We then find
\begin{align}
G\S=ig^{s}G\ell Dg+ig^{s}(\widetilde{L}^{\rm SEX}-\ell)Dg
=g(1-L^{\rm HSEX} v)\widetilde{L}^{\rm SEX}g=igDL^{\rm HSEX}g.
\end{align}
Following the same steps as in Section~\ref{cohxsec}
the equation of motion for the exciton polarization becomes
\begin{align}
\frac{d}{dt}\r_{\l}=-iE_{\l \bz}\r_{\l}
-i\W_{\l}- \frac{1}{2}\sum_{\l'\blQ}\widetilde{\G}^{{\rm pol}}_{\l\l'\blQ}(t)
\r_{\l}(t),
\label{drldt3}
\end{align}
where the polarization rates are expressed in terms of excitonic 
energies and wavefunctions
\begin{align}
\widetilde{\G}^{\rm pol}_{\l\l'\blQ}&=
2\p\sum_{\a}\frac{|\callG^{b,\l'\l}_{\a}(\blQ)|^{2}}{2\w_{\a\blQ}}
\Big[\d(E_{\l'\blQ}-E_{\l\bz}+\w_{\a\blQ})(1+f^{\rm ph}_{\a-\blQ})+
\d(E_{\l'\blQ}-E_{\l\bz}-\w_{\a\blQ})f^{\rm ph}_{\a\blQ}\Big],
\label{gammabare}
\end{align}
and $\callG^{b}$ is the {\em bare} exciton-phonon couplings
\begin{align}
\callG^{b,\l\l'}_{\a}(\blQ)\equiv 
\sum_{c_{1}c_{2}v_{1}\blk_{1}}
A^{\l\blQ\ast}_{c_{1}v_{1}\blk_{1}}g^{c_{1}c_{2}}_{\a-\blQ}(\blk_{1}+\blQ)
A^{\l'\bz}_{c_{2}v_{1}\blk_{1}}
-
\sum_{c_{1}v_{1}v_{2}\blk_{1}}A^{\l\blQ\ast}_{c_{1}v_{1}\blk_{1}}
g^{v_{2}v_{1}}_{\a-\blQ}(\blk_{1}+\blQ)A^{\l'\bz}_{c_{1}v_{2}\blk_{1}+\blQ}.
\label{callGx2irrbare}
\end{align}

The equation of motion Eq.~(\ref{dnincdt3}) for the incoherent 
exciton numbers can also be reformulated in terms of only excitons 
and bare $e$-$ph$ couplings. 
According to Eq.~(\ref{altL0}) we have $L=L^{(v)}-iL^{(v)}gDgL^{(v)}$.
Using this expression  and following the same 
steps as in Section~\ref{incexsec} we find
\begin{align}
\frac{d}{dt}
N^{\rm inc}_{\l\blQ}=
\sum_{\l'}\widetilde{\G}^{\rm pol}_{\l'\l\blQ}(t)\,
|\r_{\l'}(t)|^{2}.
\label{dnincdt4}
\end{align}
It is straightforward to verify that also in in this alternative 
formulation the total number of excitons satisfies Eq.~(\ref{dNtotdt}).

The above formulation overlooks an important detail. 
For a semiconductor in equilibrium at low temperature it is 
reasonable to set to zero all matrix elements of $L$ except for 
$L_{cv,c'v'}$ and $L_{vc,v'c'}$. From Eq.~(\ref{gscreened}) we have (omitting the 
dependence on momenta and times)
\begin{align}
&g^{s,cc'}=g^{cc'}-i \sum_{\a\bar{\a}\b\bar{\b}}
v_{c\bar{\a} \a c'}L_{\a\bar{\a},\b\bar{\b}}g^{\b\bar{\b}}.
\end{align}
We see that the dressing of 
$g^{cc}$ is due to the interband bare $e$-$ph$ couplings 
$g^{cv}$ and $g^{vc}$. The same holds true for the dressing of $g^{vv'}$.
Since the interband couplings are not negligible, they 
must be included. This complicates the whole theory as the equations 
of motion can no longer be closed solely on exciton numbers.      

\end{appendix}


\nolinenumbers

\end{document}